%% file: main.tex
\documentclass[a4paper,11pt]{article}

\pdfoutput=1
\usepackage{./style_files/jheppub} 
\usepackage{bm}
\usepackage{latexsym}
\usepackage{amsmath}
\usepackage{epsfig}
\usepackage[T1]{fontenc} 
\usepackage{textcomp}
\usepackage{amssymb}
\usepackage{graphicx}
\usepackage[section]{placeins}
\graphicspath{ {./images/} }
\usepackage{titlesec}
\setcitestyle{square}
\usepackage{multirow}
\usepackage{caption}
\usepackage{booktabs}
\DeclareMathOperator{\sech}{sech}
\DeclareMathOperator{\cosech}{cosech}
\DeclareMathOperator{\sign}{sign}
\usepackage[page,toc,titletoc,title]{appendix}
\usepackage{tocloft}
\usepackage{blindtext}
\newcommand\numberthis{\addtocounter{equation}{1}\tag{\theequation}}
\usepackage{comment}
\usepackage{soul}

\title{\boldmath Wigner distributions in Rindler spacetime and nonvacuum Minkowski states}

\author[a,b]{Nitesh K. Dubey}
\author[a]{Sanved Kolekar}

\affiliation[a]{Indian Institute of Astrophysics \\ Block 2, 100 Feet Road, Koramangala, Bengaluru 560034, India.}
\affiliation[b]{Pondicherry University, R.V.Nagar, Kalapet, Puducherry-605014, India}

\emailAdd{nitesh.dubey@iiap.res.in}
\emailAdd{sanved.kolekar@iiap.res.in}

\abstract{In the 1970s, Fulling, Davies, and Unruh demonstrated that the vacuum state perceived by an inertial observer in Minkowski space appears as a thermal bath to a uniformly accelerated observer. We explore the transformation of the Wigner distribution of a real scalar field from an inertial to a Rindler frame, utilizing both Minkowski and Unruh modes. We present a general expression for the reduced Wigner distribution for a specific set of massless scalar field configurations, and validate it against known distributions within this set. This includes arbitrary Gaussian states of Unruh-Minkowski modes, the Minkowski vacuum state, the Rindler vacuum, and the thermal bath of Unruh particles. Additionally, we analyze several other distributions, such as a peaked frequency distribution, a slight deviation from the Minkowski vacuum, and a distribution with a Fermionic component in the Rindler frame. The conclusions are discussed.}

\include{./style_files/packages}
\include{./style_files/PSI_packages}

\begin{document}
\maketitle
\flushbottom

\newpage

\section{Introduction}
The fundamental advancement of quantum field theory in curved space-time has been significantly influenced by the concept of how a system's state, which associates an operator with its expectation value in a linear manner, depends on reference frames. A noteworthy application of this theory arises when we restrict to a specific wedge, denoted as R, within Minkowski space-time: R:= $\{ x \in \mathbb{R}^{1,3} | x_1 > |x_0|  \}$. This wedge, referred to as Rindler spacetime is entangled with the complementary left wedge and their union constitutes a globally hyperbolic space-time. In Minkowski spacetime, the existence of a unique Poincaré-invariant state, known as the Minkowski vacuum, is firmly established. Various global families of inertial observers in Minkowski spacetime converge on a common understanding of particle content within any given field state \cite{Chmielowski:1994gb}. However, the one-parameter group of Lorentz boost isometries can be used to construct a Rindler spacetime. In the literature, Fulling \cite{PhysRevD.7.2850, Fulling} first gave the Bogolyubov transformation of creation and annihilation operators from the inertial frame to the Rindler frame. The interpretation of particle content obtained by Fulling using Bogolyubov transformation as a black body spectrum was later given by Davies \cite{Davies:1974th}. However, the correct interpretation of these results was later developed by Unruh \cite{Unruh:1976db}. It was shown that the vacuum state of an inertial observer appears as a thermal bath for the Rindler observer. Subsequent advancements led to the development of various versions of thermalization theorems \cite{10.1143/PTP.88.1}. In the literature, the behavior of the thermal bath of Unruh particles in the inertial frame as seen by a Rindler observer \cite{Costa:1994yx, Kolekar:2013aka} has also been discussed. There are also attempts to understand the Rindler vacuum \cite{Rajeev:2019bzv}. The role of Unruh effect in triggering QCD phase transitions has been explored in \cite{QCD}. The Unruh effect is thought to be the most spectacular manifestation of the vacuum entanglement. Recent advancements in our understanding of quantum entanglement offer a unique avenue for scrutinizing the quantum nature of gravity, as indicated by a study \cite{Bose:2017nin}. Furthermore, in the context of stochastic quantum field theory in gravity, entanglement can be conveyed through the local Unruh effect without necessitating the presence of gravitons \cite{PhysRevD.107.056014}. This, in turn, encourages us to delve deeper into the intricacies of the Unruh effect.

In field theory description, the vacuum state is characterized by being both pure and Gaussian. The vacuum for a given observer depends upon the eigenfunctions/ eigenmodes of the field operators defined with respect to the proper time of that observer. Given the direct interaction between an accelerated detector and Rindler modes, these modes become the more suitable choice for quantizing the field from the perspective of accelerated observers. Moreover, for inertial observers, the Minkowski modes are a natural choice for quantizing the field \cite{Costa:1994yx}. Nevertheless, one can also design an interaction Hamiltonian in a way that it couples with Unruh modes. It's important to highlight that many of the conclusions drawn from the Unruh Fock space can be extended to the original Minkowski plane wave Fock space, albeit requiring a basis transformation from plane waves to wave functions defined in log x space. This dual approach encourages one to leverage both Minkowski and Unruh modes in the comprehensive exploration of a wide range of states.

In physical situations, such as the evolution of the Higgs field in the early universe, models of galaxy formation, and fields around astrophysical objects, are not in a vacuum or thermal equilibrium state \cite{Horn:2020wif, PhysRevD.52.6770, Mazur:2012da}. 
One also ends up dealing with non-vacuum states while trying to observe the Unruh effect through analogue models in a laboratory \cite{Tallent:2024okw, Bunney:2023vyj, PhysRevLett.106.021302, Weinfurtner:2013zfa, PhysRevD.92.024043, PhysRevLett.114.036402, PhysRevLett.105.203901, PhysRevA.90.033607, PhysRevLett.85.4643, CarlosBarceló_2001, PhysRevA.78.021603, PhysRevA.80.043603}. In particular, \cite{PhysRevLett.114.036402} reported a study of superfluid hydrodynamics of polaritons along a one-dimensional microstructure for the detection of analogue Hawking radiation. \cite{PhysRevLett.105.203901} discussed the experimental realization of gravitational analogue using ultrashort laser pulse filaments. Furthermore, \cite{PhysRevA.90.033607, PhysRevLett.85.4643, CarlosBarceló_2001, PhysRevA.78.021603, PhysRevA.80.043603} studied theoretically the analogue Hawking radiation in Bose-Einstein condensates. There are several other cases wherein non-vacuum states are relevant. For example,  states of low energy states in FLRW spacetimes has been introduced in \cite{Olbermann:2007gn, Them:2013uka}. \cite{Buchholz:2001qj, Solveen:2010fn} discussed the notion of nonequilibrium and local thermal equilibrium states. Later \cite{Buchholz:2003cz} applied the concept of local thermal equilibrium states to the space-time evolution of a hot bang at the origin of Minkowski space. It has also been claimed that the KMS property does not entail local thermal equilibrium \cite{Solveen:2012ai}. Further in the case of the retrieval of shock wave memory of spacetime \cite{Majhi:2020pps}. In most of these cases, it is the relativity of states and their particle content interpretation as seen from different frames which forms the crucial aspect in all these investigations. This, in turn, motivates us to investigate the non-vacuum states from perspectives of different observers, particularly in the context of the fundamental phenomenon such as the Unruh effect.  

The Rindler basis sees a monochromatic wave of Minkowski modes as a non-chromatic superposition of all frequencies. Due to this mode mixing property, exploring beyond the single/double Minkowski particle state in the Rindler frame becomes complicated analytically, and indeed, the literature is scarce in this context \cite{Falcone:2022jbj, PhysRevD.106.045013}. Below, we highlight some of the special cases that were explored earlier. The response of an accelerated detector in de Sitter spacetime was discussed in \cite{Deser:1997ri}, where the effective temperature observed is found to be the Pythagorean sum of the de Sitter temperature and the acceleration temperature. \cite{PhysRevD.89.044036} investigated the uniformly accelerated observer in a thermal bath of Unruh particles in Minkowski spacetime. The authors of \cite{Kolekar:2013aka} studied the transformation of a thermal bath of Unruh particles from an inertial frame to the Rindler frame using the density matrix formalism and concluded an interesting indistinguishability of thermal and quantum fluctuations. 
While \cite{Kolekar:2013aka} assumed thermality in only one of the Unruh modes. It would be interesting to explore the possibility of considering the state to be thermal in both Unruh-Minkowski modes that yield the same Rindler frequency and, more importantly, analyze the corresponding state in terms of Minkowski modes. Furthermore, (i) exploring a Minkowski state which is close to vacuum but not exactly the vacuum --- this deviation could correspond to a genuine source in the Minkowski frame or arise as some noise in the system which is difficult to remove (ii)  a uniform frequency distribution state in Minkowski frame corresponding to a band filter which would allow only certain frequencies to pass through (iii) interplay between bosonic and fermionic statistics ---  all such states to analyze from the perspective of the Rindler observer is interesting and such an investigation may form a basis to explore more complicated states. 

In the present manuscript, we employ the Wigner functional formalism \cite{Mrowczynski:1994nf, PhysRevD.87.065026, Cembranos:2021sfj, Padmanabhan_2010}, and provide a relatively simpler and tractable treatment to investigate many non-vacuum Minkowski states. In many situations, it is possible to associate a unique density matrix with any state of a system. The knowledge of density matrix can be used to compute different correlations. It can also be used to find out entropy and, hence, thermodynamic properties of the system. There are many techniques in quantum optics that have been used in quantum field theory in curved space \cite{Scully:2017utk, PhysRevLett.121.071301}. The Wigner functional, mainly used in quantum optics, is a phase-space representation of the density matrix, which simplifies calculations. The Wigner distribution has also been used to transform the inertial-frame vacuum state of a scalar field to the Rindler frame \cite{Ben-Benjamin_2020}.

In section 2, we provide a brief overview of the Wigner functional framework for the massless real scalar quantum fields in flat spacetime. We then provide a general transformation of the Wigner distribution of a massless scalar field in Minkowski spacetime to that of the Rindler spacetime for a particular subset, yet broadly general, which encompasses a large set of states of Minkowski and Unruh-Minkowski modes including the non-vacuum states motivated above. This subset is more formally defined in subsection 2.2 and the detailed calculation of the transformation is presented in Appendix A. Additionally, we highlight several general features of the reduced Wigner functional for these states. We also provide a general expression for the expectation value of the number density operator for each of these Wigner functionals in subsection 2.4. In section 3, we first verify the general expression obtained for known cases such as the vacuum Minkowski state and Rindler vacuum for consistency, while also highlight some new findings in the latter case. We then proceed to discuss several new distributions namely (i) a Minkowski state which is close to vacuum but not exactly the vacuum (ii) a peaked frequency distribution state in Minkowski frame (iii) a thermal bath in both Unruh-Minkowski modes (iv) a state which offers interplay between bosonic and fermionic statistics and their corresponding reduced Wigner distributions in the Rindler frame along-with expectation values of number density in each. We summarize our results in the discussion section \ref{discussion}. Use is made of natural units, namely $\hbar$=c= $k_B $=1, throughout the paper.

\section{Vacuum and non-vacuum Minkowski states in Rindler frame}

We begin by introducing the Wigner functional framework particularly for the massless real scalar quantum fields in flat spacetime.   

\subsection{Wigner Functional ---Setup}

One can describe physical systems in terms of density function operator $\rho$, which is a non-negative hermitian operator of trace unity acting on a Hilbert space $\mathcal{H}$ \cite{Harlow:2014yka}. The components of $\rho$ expanded in a complete set of basis eigenstates constitute the density matrix. There exist numerous scenarios, one such example being coherent states of photons, where the uncertainty relation is minimized. It makes the coherent state to be concentrated along classical trajectories. In this paper, we use the basis of coherent states to describe the quantum states. The phase-space representation of the density matrix, which is a pseudo-probability distribution, is known as the Wigner distribution \cite{scully_zubairy_1997}. One uses c-number correspondence to get a c-number function corresponding to any operator \cite{scully_zubairy_1997}. The Wigner functional approach serves the closest analog to classical physics, having roots in quantum theory. One can refer to  \cite{Mrowczynski:1994nf, PhysRevD.87.065026, Cembranos:2021sfj} for a detailed study of the Wigner functional approach. It can be said that this feature makes the classical field limit more transparent compared to other formulations. For a quantum harmonic system, one can express the Wigner distribution in the following form for the $n^{th}$ number state of the harmonic oscillator \cite{Ben-Benjamin_2020}.
\begin{equation} \label{eq:1}
    W^{(n)}(\alpha) = \langle n| 2 e^{-2 (\hat{a}^ {\dagger}- \alpha ^*);(\hat{a} - \alpha)} |n \rangle = \frac{1}{n!} \langle 0| (\hat{a})^n 2 e^{-2 (\hat{a}^ {\dagger}- \alpha ^*);(\hat{a} - \alpha)} (\hat{a} ^ {\dagger} )^n |0\rangle
\end{equation}
Here $`;$' denotes Schwinger operator ordering,  $\hat{a}$ and $\hat{a} ^\dagger$ denote the annihilation and creation operators, and $\alpha, \alpha ^*$ $\in \mathbf{C}$ with $\alpha$ being eigenvalue of annihilation operator. The above expression can be formally thought of as a derivative acting on vacuum Wigner distribution. One can also think of the Wigner distribution as the expectation value of the displaced parity operator. 

We consider a system of massless real scalar field in the background Minkowski spacetime and apply the Wigner functional approach to analyze how the functional transforms between inertial and Rindler bases states. It is, in general, known how to transform Wigner distribution from one set of complete bases to a new one, corresponding to different reference frames of different observers. We will be dealing with the Wigner distribution based on coherent states \cite{Ben-Benjamin_2020} because it practically simplifies the relevant calculations. In general, one could adopt the Wigner functional approach to explain the dynamics of almost all quantum field systems where the notion of a density matrix can be suitably defined \cite{Mrowczynski:1994nf}. We begin by expressing the massless real scalar field in terms of the basis of the Minkowski plane wave mode solution of the Klein-Gordon equation, as follows: \\
\begin{equation}  \label{eq:2}
\hat{\phi}(t,\mathbf{x}) = \iiint _V \frac{d^3 \mathbf{k} }{(2 \pi)^{3/2}\sqrt{2 \omega _{\mathbf{k}}}} [\hat{a}_{\mathbf{k}} e^{ikx}  + \hat{a}^\dag _{\mathbf{k}} e^{-ikx} ].
\end{equation}
\\
Here $\hat{a}_{\mathbf{k}}$, $\hat{a}^\dag _{\mathbf{k}}$ are standard annihilation and creation operators corresponding to the Minkowski mode  $\mathbf{k}$, and $\{t, \mathbf{x} \}$ are standard Minkowski coordinates. The above expansion is not formally convergent and lacks the notion of being an observable associated with a spacetime point; however, one can interpret of it as an operator-valued distribution \cite{Wald:1995yp}. In the momentum representation, the quantum field is essentially a collection of simple harmonic oscillators, which allows one to apply the Wigner function formalism as described in Eq.\eqref{eq:1} for each mode $\mathbf{k}$. For the sake of simplicity and analytical tractability, we work in (1+1) dimensions for the rest of the paper. We begin by choosing a particular form for the Minkowski-Wigner distribution function, which represents a wide set of initial quantum states. The motivation for such a choice comes from the fact that the Wigner distribution of a thermal bath and Gaussian quantum states, which play a significant role in several areas of theoretical and experimental physics, form a subset of the form chosen. We consider the form of the Wigner distribution in the Minkowski basis to be \\
\begin{equation} \label{eq:3}
W_M = \bar{N} \exp(-2 \int _{-\infty} ^ {+\infty} \int _{-\infty} ^ {+\infty}\frac{dk dk'}{4 \pi^2} a_{k'} ^* a_{k} f(k,k')),
\end{equation}
\\
where the subscript $M$ represents an inertial frame, $\bar{N}$ is the normalization factor, $a_{k'} ^* $ and  $a_k$ are c-numbers corresponding to creation and annihilation operators in an inertial frame, obtained by Wigner-Weyl correspondence \cite{Ben-Benjamin_2020}. One can refer to  \cite{Mrowczynski:1994nf} for a discussion of the Wigner functional for thermal equilibrium states in the field representation. The function $ f(k,k')$ is a two-point function, i.e., we allow the Wignar functional to have off-diagonal elements in the k-k' space. This allows us to consider Minkowski states beyond the popular $ f(k,k') $ = $ \delta(k-k') f(k)$ class of states usually considered in the literature. The Minkowski vacuum state and the thermal bath both fall in the latter set of states. In general, the two-point function $f(k,k')$ can be any Hermitian positive two-point function of Schwarz space, or it should have compact support. 

To transform the Wigner functional in terms of the basis of charts of the accelerated frame of reference, we first define the Rindler transformation by considering an observer moving with a uniform linear acceleration `a', along the x-axis with trajectory taken as $(\frac{1}{a} \sinh{a \tau}, \frac{1}{a} \cosh{a \tau} )$. Here, $\tau$ represents the observer's proper time, and the motion is restricted to the t-x plane. The wedge part of the entire Minkowski spacetime accessible to the uniformly accelerated observer defined above constitutes a globally hyperbolic spacetime along with the union with its complementary left wedge, though not geodesically complete, is called the right Rindler wedge/patch or the Rindler spacetime. The global hyperbolicity allows one to quantize the field and define modes restricted to the right Rindler wedge. However, the vacuum of an inertial observer, i.e., the Minkowski vacuum, is defined over the full Minkowski spacetime. The real massless scalar field can be expanded in terms of the Rindler plane wave basis mode solutions  of the Klein-Gordon equation as 
\begin{equation}  \label{eq:4}
\hat{\phi}(T,X) = \int _{-\infty} ^{+\infty} \frac{d K }{(2 \pi)^{1/2}\sqrt{2 \omega _{K}}} [\hat{b}_{{K,R}} e^{i K_\mu X^\mu}  + \hat{b}^\dag _{K,R} e^{-i K_\mu X^\mu} ] + [R \leftrightarrow L ],
\end{equation}
where $|K|$ is the Rindler frequency, {$T, X$} are Rindler coordinates and $\{  \hat{b}_{{K, R}}, \hat{b}^\dag _{K, R},\hat{b}_{{K, L}}, \hat{b}^\dag _{K, L} \}$ are creation and annihilation operators corresponding to left and right Rindler wedges. This set of creation and annihilation operators is related to the Minkowski creation and annihilation operators by Bogolyubov transformations.

In the next two sub-sections, we provide a general transformation of the Wigner distribution of a massless scalar field in Minkowski spacetime to that of the Rindler spacetime for a particular subset, yet broadly general, which encompasses a large set of states of Minkowski and Unruh-Minkowski modes including the non-vacuum states motivated in the introduction.

\subsection{Reduced Wigner Functional for Inertial Minkowski Particles }

One would like to transform the Minkowski Wigner functional defined in Eq.(\ref{eq:3}) with the help of Bogoliubov transformations. However, it is well known that for a general two-point function $ f(k,k')$, the procedure is analytically untractable due to the mode mixing property of the Rindler transformations for Minkowski modes. 
However, for the set of states we would like to investigate such as the (i) exploring a Minkowski state which is close to vacuum but not exactly the vacuum - this deviation could correspond to a genuine source in the Minkowski frame or arise as some noise in the system which is difficult to remove (ii)  a uniform frequency distribution state in Minkowski frame corresponding to a band filter which would allow only certain frequncies to pass through (iii) A  thermal state in both Unruh-Minkowski modes that yields an interesting  indistinguishability of thermal and quantum fluctuations in the Rindler frame and more importantly analyze the corresponding state in terms of Minkowski modes (iv) a state which offers interplay between bosonic and fermionic statistics - it turns out that one can still get around the mode mixing complication and calculate the the reduced Rindler Wigner functional for inertial frame Minkowski mode states by selecting a specific set of two-point weight functions $ f(k,k')$ which we shall describe below. Such a choice does not limit one to a narrow set of non-vacuum states that one can explore, but in fact, it is still quite general and allows one to explore a wide set of states.

For our purpose stated above, it suffices to choose a subset of $ f(k,k')$ such that the density functional in the right Rindler basis modes becomes diagonal in the $K-K'$ space. This is realized if we consider those Hermitian positive definite two-point functions whose 2D Fourier transform with respect to $(K/a, K'/a)$ after a redefinition of the modes as  $k=e^t, k'=e^{t'}$ and weighted by $e^{(t+t')/2}$  is diagonal for some continuous variable $K$ and positive nonzero constant `a'. The same can be considered as a 2D Mellin transform which is required to be diagonalized. Such a requirement on $ f(k,k')$ can be formally written for all 4 quadrants of the $k-k'$ space in a matrix sense as
\begin{equation} \label{eq:5}
\int _{0} ^{+\infty} \int _{0} ^{+\infty} \frac{dk dk'}{\sqrt{|k||k'|}} f(k,k') |k'/a|^{-iK'/a} |k/a|^{+iK/a} = g_1(K,K') \delta(K-K')
\end{equation}

\begin{equation} \label{eq:6}
\int _{0} ^{+\infty} \int _{0} ^{+\infty} \frac{dk dk'}{\sqrt{|k||k'|}} f(k,-k') |k'/a|^{-iK'/a} |k/a|^{+iK/a} = g_2(K,K') \delta(K-K')
\end{equation}

\begin{equation} \label{eq:7}
\int _{0} ^{+\infty} \int _{0} ^{+\infty} \frac{dk dk'}{\sqrt{|k||k'|}} f(-k,k') |k'/a|^{-iK'/a} |k/a|^{+iK/a} = g_3(K,K') \delta(K-K')
\end{equation}

\begin{equation} \label{eq:8}
\int _{0} ^{+\infty} \int _{0} ^{+\infty} \frac{dk dk'}{\sqrt{|k||k'|}} f(-k,-k') |k'/a|^{-iK'/a} |k/a|^{+iK/a} = g_4(K,K') \delta(K-K')
\end{equation}
for some real two point functions $g_1(K,K')$, $g_2(K,K')$, $g_3(K,K')$  and $g_4(K,K')$. We denote the diagonal elements $g_s$(K,K) as $g_s$(K). In principle, the above equations can also be inverted using the inverse Fourier transform to find \( f(k, k') \) for a given set of \( g_i(K, K') \), provided the inverse transformation exists, and the obtained \( f(K, K') \) is well-defined, so it can be considered as a weight function of the Wigner functional. The inversion property allows us to first choose a Rindler density functional and then obtain a corresponding Minkowski density functional, as we shall demonstrate in the next section. However, one should note that the Minkowski state obtained in this way is not unique because the degrees of freedom in the left wedge have been traced out to get the reduced Wigner functional in the right Rindler wedge. There can be other Minkowski states with a weight function $f(k,k')$ that does not belong to the subset of functions satisfying properties Eqs.(\ref{eq:5}-\ref{eq:8}) and yet may yield the same Rindler state. This behavior could be expected because each Minkowski mode has been mixed up with modes of all frequencies during the Rindler transformation.  However, if we have a Minkowski space Wigner functional that belongs to the above mentioned subset, then the Rindler space Wigner functional will be unique. Given the Wigner distribution, physical quantities of interest, such as expectation value of number density, entropy, correlation, .. etc., can be easily determined. The degrees of freedom in the Wigner functional in both inertial and accelerated frames are related by Bogoliubov transformation \cite{Falcone:2022jbj}. The degrees of freedom in the left and right Rindler wedges are entangled. Since the left Rindler wedge is inaccessible to the uniformly accelerated observer in the right wedge, we need to trace over unobserved degrees of freedom after applying the Bogoliubov transformation, which is the standard procedure to obtain the reduced Wigner functional for the right Rindler observer. We provide a detailed derivation of the reduced Wigner distribution in Appendix A. The reduced Wigner distribution corresponding to the Minkowski Wigner functional in Eq.(\ref{eq:3}) and for $f(k,k')$ satisfying Eqs. (\ref{eq:5}-\ref{eq:8}) is obtained to be
\begin{multline} \label{eq:9}
    W_{R_{\text{reduced}}} = N \exp \left[ \frac{-1}{8 \pi ^4 a ^2} \int _0 ^{\infty} dK \, \left| K \right| \left| \Gamma \left( \frac{iK}{a} \right) \right| ^2 \left( J(K) \left| b_R (K) \right| ^2  + R_1(K) b_R ^* (K) b_R ^* (-K) \right. \right. \\
    \left. \left. + R_2 (K) b_R(K) b_R (-K) + L \left| b_R (-K) \right|^2 \right) \right] \\
   \hphantom{W_{R_{\text{reduced}}}} = N \exp \left[ \frac{-1}{8 \pi ^3 a} \int _0 ^{\infty} \frac{dK}{\sinh{\pi |K|/a}} \left( J(K) \left| b_R (K) \right| ^2  + R_1(K) b_R ^* (K) b_R ^* (-K) \right. \right. \\
    \left. \left. + R_2 (K) b_R(K) b_R (-K) + L \left| b_R (-K) \right|^2 \right) \right],
\end{multline}
which is diagonal in $K-K'$ space. The explicit expressions for weight functions J(K), $R_1 (K)$, $R_2(K)$, and L(K) are given in Eqs. \eqref{eq:49} to \eqref{eq:51} of Appendix A.

Some interesting technical features can be deduced for these weight functions and $ W_{R_{\text{reduced}}}$ which we shall highlight below. \\
\\
(a) $ g_2(K,K) \leftrightarrow g_3(K,K) \Rightarrow R_1(K) \leftrightarrow R_2(K) $ where the doublearrow represents an interchange operation\\
(b) If  $ g_2(K,K) = g_3(K,K) = 0 $ then $R_1(K)=R_2 (K)=0$\\
(c) L(-K) = L(K) , J(-K) = J(K) , $R_1(-K)=R_2 (K)$, and $R_2(-K)=R_1 (K)$\\
(d) If $ g_2(K,K) = g_3(K,K) $ =  $g_2(-K,-K) = g_3(-K,-K) $ and $ g_1(K,K) = g_4(K,K) $ = $g_1(-K,-K) = g_4(-K,-K)$ $\Rightarrow J(K)=L(K)$ \\
\\
The first property suggests that if the weight function $f(k, k')$ of Minkowski Wigner functional satisfies $f(k, -k')$ = $f(-k, k')$, then the weights $g_2(K, K')$ and $g_3(K, K')$ are also equal. As a result weights $R_1(K) $ and $R_2(K)$ in the reduced Rindler Wigner functional, which are linked to the cross terms of positive and negative frequencies, are the same. In those cases where they are not equal, swapping them results in a corresponding interchange of weights, i.e., $R_1(K) \leftrightarrow R_2(K)$ within the Rindler wedge. The second property asserts that if $g_2(K, K)$ and $g_3(K, K)$ vanish, then the corresponding cross term $R_1(K)$ and $R_2(K)$ in the Rindler frame also vanish. As a consequence, the anomalous averages such as $\langle b_R^* (K) b_R^* (-K) \rangle$ in Eq.(\ref{eq:9}) vanish. Non-zero anomalous averages play an important role in several phenomena, such as the  BCS theory of superconductors \cite{landau1980course, Akhmedov:2021vfs, mattuck1992guide,zubarev1996statistical}. The final property elucidates that if the weight of Minkowski Wigner functional exhibits reflection symmetry about the origin, i.e., $f(k,-k')$ = $f(-k,k')$ and $f(k,k')$ = $f(-k,-k')$  then from Eq. \eqref{eq:5} to Eq.\eqref{eq:8} we have, $ g_2(K,K) = g_3(K,K) $ =  $g_2(-K,-K) = g_3(-K,-K) $ and $ g_1(K,K) = g_4(K,K) $ = $g_1(-K,-K) = g_4(-K,-K)$ --- which after substituting in Eq.\eqref{eq:14} give an equal number of particles and anti-particles in the Rindler frame.

In this subsection, we have thus provided a general expression for the reduced Wigner distribution in the Rindler frame starting from a Wigner distribution in the Minkowski frame corresponding to our subset of cases of interest. In the next subsection, we shall do so for the Unruh-Minkowski modes as well.

\subsection{Wigner Functional for UM modes}
While the concept of particles is not a prerequisite for comparing the relationship between the Fock basis and Bogoliubov transformations in two frames, specific sets of modes and particle states naturally manifest as a consequence of the symmetries inherent in the spacetime under consideration. One such set of modes is the Unruh Minkowski (UM) modes which is a linear superposition of positive frequency Minkowski modes. Its ground state coincides with the Minkowski vacuum. However, other excited states are different. 
It turns out that a subset of the reduced Rindler frame Wigner distribution in Eq.\eqref{eq:9}, when $R_1(K)$ = $R_2(K)$ = 0 and J(K) =  L(K), can be obtained by a very specific set of inertial frame Wigner functionals in terms of UM modes. This can be seen as follows. It is known that there are two UM modes of positive frequency corresponding to each Rindler frequency. One can express the Wigner distribution in terms of modes corresponding to these UM particles as
\begin{equation}  \label{eq:10}
    W_{UM} = \prod_{\xi >0} N_\xi e^{-  p(\xi) |\alpha_{\xi}|^2 - q(\xi) |\alpha_{-\xi}|^2  -  r(\xi) \alpha_{\xi} \alpha_{-\xi} - s(\xi) \alpha_{\xi} ^* \alpha_{-\xi} ^*  }
\end{equation}
Here, $\alpha $ and $\alpha^*$ are c- number representations of annihilation and creation operators for Unruh-Minkowski particles. The functions $p(\xi)$ and $q(\xi)$, which are positive, nonzero, real, and smooth, dictate the weights linked to $\alpha_{\xi}$ and $\alpha_{-\xi}$. Similarly, $r(\xi)$ and $s(\xi)$ are real smooth functions. An additional constraint arises from the requirement for the expectation value of number density to be finite, expressed as $pq \neq rs$. We emphasize that one does not require the set of conditions in Eqs. \eqref{eq:5}-\eqref{eq:8} when dealing with the Unruh-Minkowski modes here. One can also consider a more general form of the weight functions that depend on two arguments, for example, p($\xi$, $\xi '$), such that the density functional is off-diagonal in $\xi-\xi'$ space as in Eq.\eqref{eq:3}. Here, we consider only the diagonal form as given in Eq.\eqref{eq:10} since it suffices to explore different cases in the next section. To obtain the reduced Wigner distribution in the right Rindler frame, we proceed in a manner similar to that in the above section to obtain
\begin{equation} \label{eq:11}
    W_{R_{reduced}} = \prod_{K >0} N_K \exp \left\{-   \frac{ 2 ( p q - r s) \sinh (\frac{\pi K}{a})  }{ p e^{-\pi K/a }  + q e^{\pi K/a } - r - s }  (|b_{R}(K)|^2 + |b_{R}(- K)|^2 )\right\}
\end{equation}
\\
where, $ p \equiv p(K) $ , $  q \equiv p(K)  $ , $  r \equiv r(K)  $ and $  s \equiv s(K)  $ are the same weight functions $p(\xi) $, $q(\xi)$, $r(\xi)$ and $s(\xi)$ in Eq.\eqref{eq:10} with the arguments replaced by $|\xi| $= $\frac{2 \pi }{a}$. 
\subsection{Number Density in momentum space}

Having obtained the Wigner distribution in two distinct frames, one can now proceed to compute the expectation values of any operator in both frames, especially the expectation value of the number density operator. In this subsection, we provide an expression for the expectation value of the number density associated with various established mode sets and defined by
\begin{equation} \label{eq:12}
    \langle n \rangle \equiv \int d^2 \alpha (\alpha ^* \alpha -1/2) W
\end{equation}
where integration runs over the respective $\alpha _s$ in their own phase space corresponding to each W$(\alpha,\alpha^*)$. Substituting our Wigner distribution,$W_M$, and performing the Gaussian integration, we get the following expressions for the particle number density expectation value for Minkowski modes as
\begin{eqnarray} \label{eq:13}
    \langle n_M(k) \rangle _i  = \frac{1}{2} \left[\left(\frac{f(k,k')}{2 \pi}\right)^{-1} _{ii} - \delta(0) \right].
\end{eqnarray}
Here, the subscript ii in the expression of $ \langle n_M(k) \rangle_i$ refers to the $ii^{th}$ diagonal element in the inverse matrix. The $\delta(0)$ term in the above equation comes due to the continuum choice of modes. This divergence can be regularized by considering a box of finite size and dividing both sides by its volume. We denote the regularized expectation values by a tilde over $\langle n_s \rangle$, i.e., $\langle \tilde{n}_s \rangle$. The regularized expectation values for the UM modes, by using  $W_{UM}$ are obtained as \footnote{To get the inverse of the diagonal matrix, say $ A(k,k)$, let us write it's functional form as $ A(k,k') \delta(k-k')$. Let  $ B(k,k')$ denoted its inverse then it satisfies
\begin{align*}
    \int A(k,k') \delta(k-k') B(k',k'') dk' &= \delta(k-k'')\\
    \implies  A(k,k) B(k,k'') &= \delta(k-k'')
\end{align*}
This brings a $\delta(0)$ in the expectation values of UM and Rindler particles, say $\langle n_s \rangle$. The tilde represents the regularized quantities.}
\begin{eqnarray} \label{eq:14}
    \langle \tilde{n}_{\alpha_{\xi}} \rangle =  \bigg[\bigg(p(\xi) - \frac{r(\xi) s(\xi)}{q(\xi)} \bigg)^{-1} - \frac{1}{2} \bigg] ; \langle \tilde{n}_{\alpha_{-\xi}} \rangle =\bigg[ \bigg(q(\xi) - \frac{r(\xi) s(\xi)}{p(\xi)} \bigg)^{-1} - \frac{1}{2} \bigg] .
\end{eqnarray}
\\
We note from the above expressions of the expectation value of UM particles that if any of the cross terms r or s is 0, then $ \langle n_{\alpha_{\xi}} \rangle $ will not depend upon either $r(\xi)$, $s(\xi)$ or $q(\xi)$. An analogous statement holds true for $\langle n_{\alpha_{-\xi}}\rangle$.
One gets the following expectation value $\langle \tilde{n}_R (K) \rangle $ in the Rindler frame 
\begin{equation} \label{eq:15}
    \langle \tilde{n}_R(K) \rangle = \bigg[ (L_1(K,K) - L_2(K,K) L_4 (K,K) ^{-1} L_3 (K,K))^{-1} -\frac{1}{2} \bigg] ,
\end{equation}
where functions $L_1,L_2,L_3$ and $L_4$ are related to the weight functions in Eq.\eqref{eq:9} as \\
\\
$L_1$ = $\frac{J(K,K)}{8\pi^2 a \sinh (\pi |K|/a)}$; $L_2$ = $\frac{R_1(K,K)}{8\pi^2 a \sinh (\pi |K|/a)}$; $L_3$ = $\frac{R_2(K,K)}{8\pi^2 a \sinh (\pi |K|/a)}$; $L_4$ = $\frac{L(K,K)}{8\pi^2 a \sinh(\pi |K|/a)}.$\\

In the present section, we have thus obtained a general expression for the reduced Wigner functional and the corresponding number operator expectation value. In the next section, we shall consider and analyze several explicit examples of interest.

\section{Special cases of Wigner distributions}
There are a variety of interesting Minkowski state distributions that belong to the subset of distributions considered in the previous section. We shall investigate a few known cases to demonstrate the consistency of the Wigner functional approach with those taken previously in the literature and then proceed to discuss new ones and highlight their interesting features. 

\subsection{Minkowski Vacuum - Unruh effect}
We first consider the vacuum state of a massless scalar field in the Minkowski spacetime for an inertial observer. This is described by the well-known standard Unruh Minkowski Wigner distribution \cite{Ben-Benjamin_2020}.\\
\begin{equation} \label{eq:16}
    W_{UM} = \prod_{\xi} N_\xi e^{-2 |\alpha_{\xi}|^2  - 2 |\alpha_{-\xi}|^2 }
\end{equation}

The same can also be expressed in terms of Minkowski modes in Eq.\eqref{eq:3} by substituting the two-point function $f(k,k’)$ = 2 $ \pi \delta (k-k’) $. Both these represent the same Minkowski vacuum state. However, as explained in subsection (2.3), the transformation to a uniformly accelerated frame mixes positive and negative frequencies, and as a consequence, the uniformly accelerated observer finds the inertial frame ground state to be populated with particles. By comparing the distribution in 
Eq.\eqref{eq:16} with the form of the Wigner distribution in Unruh Minkowski form in Eq.\eqref{eq:10}, one can read off $p(\xi)$ = $ q(\xi) $ = 2  and $r(\xi)$ = $ s(\xi) $ = 0. 
Then, Eq.\eqref{eq:11} for the reduced Wigner functional in the right Rindler wedge can  be expressed as

\begin{equation}  \label{eq:17}
    W_{R_{reduced}} = \prod_{K>0} N_K e^{-2 |b_{R}(K)|^2  \tanh (\frac{\pi }{a})}
\end{equation}
This represents a thermal bath of Rindler particles with temperature T = a/$2 \pi$ and is the standard Unruh effect.
Alternatively, one could also obtain Eq.\eqref{eq:17} starting from  Eq.\eqref{eq:8} for the Minkowski Wigner functional. One gets the form of the weight functions  $ g_1(K), g_2(K), g_3 (K)$ and $g_4 (K)$ by substituting $ f(k,k’)$ = 2 $ \pi \delta (k-k’) $ in Eqs.\eqref{eq:4} to \eqref{eq:7}. We have checked that it finally leads to the same reduced Wigner functional as in Eq.\eqref{eq:17} using the expressions for different $L_s$ in Eq.\eqref{eq:15}.  The expectation value of the number density using Eq.\eqref{eq:15} then leads to the expected Planckian as\\
\begin{equation}  \nonumber
    \langle n_R(K) \rangle = \frac{\delta(0) }{e^{\frac{2 \pi |K|}{a}}-1} .
\end{equation}
It can be shown that there are no \textit{hidden correlations} in observed particles; therefore, the above  $ \langle n_R(K)\rangle $ represents a true thermal bath \cite{sciama, FENG2022136992}. One can follow \cite{Crispino:2007eb} for an elegant review of the Unruh effect.
\subsection{Rindler vacuum}

\begin{figure*}[ht!]
        \centering
        \includegraphics[width=.495\textwidth]{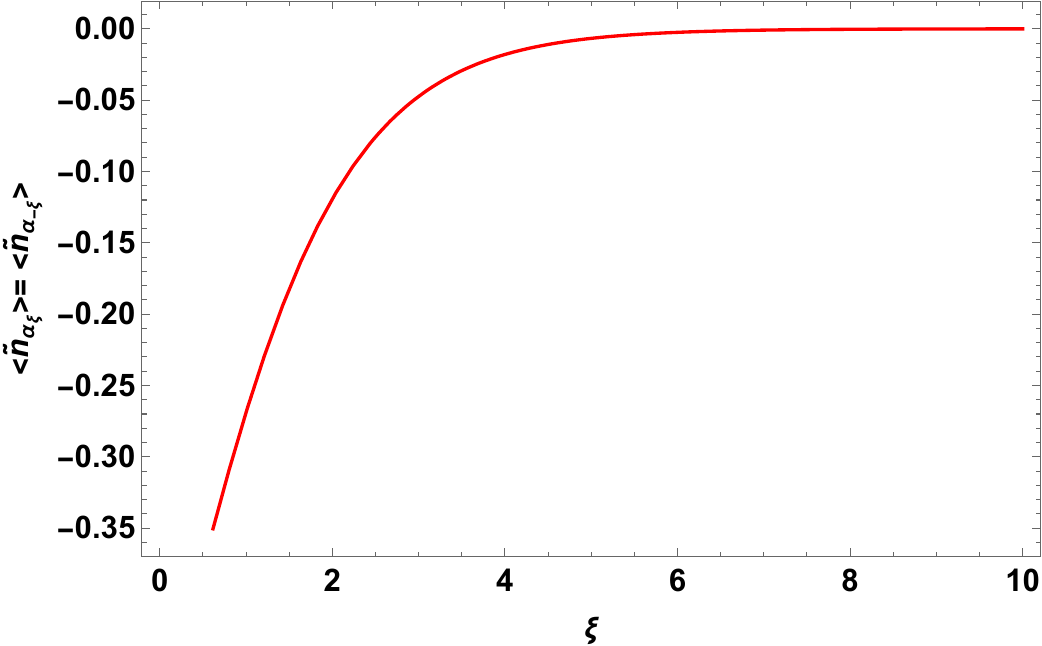}\hfill
        \includegraphics[width=.498\textwidth]{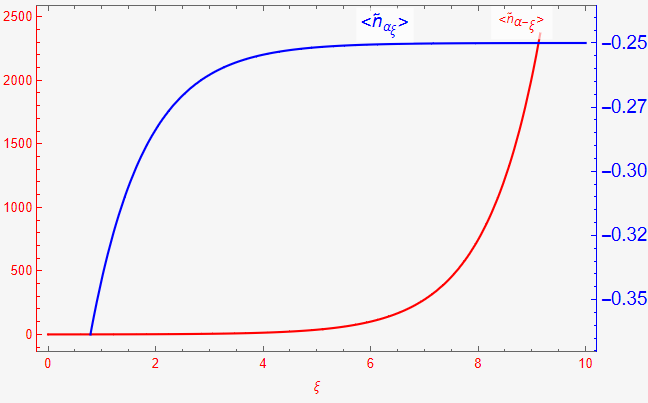}\hfill
        \caption{The left panel shows a plot of the expectation value of Unruh Minkowski particle density with choice $q(\xi)$ = $p(\xi)$, keeping r=s=0, which represents $\langle \tilde{n}_{\alpha_\xi} \rangle $= $\langle \tilde{n}_{\alpha_{-\xi}} \rangle$= $\frac{-1}{1+e^\xi}$. The right panel shows the same with $q(\xi)=e^{-\xi/2} \cosech(\xi/2)$, keeping r=s=0, which gives $\langle \tilde{n}_{\alpha_{\xi}} \rangle $= -$\frac{1+e^{-\xi}}{4}$ (Blue line) and $\langle \tilde{n}_{\alpha_{-\xi}}\rangle$ = $\frac{e^\xi-3}{4}$ (Red line).}
        \label{fig:1}
\end{figure*}

The next interesting distribution we consider is the Rindler vacuum. This is represented by a state for which the expectation value of number density in the Rindler frame is zero. The Rindler vacuum distribution can be characterized by the following reduced Wigner distribution in the Rindler frame, where the acceleration has a value greater than zero. \\
\begin{equation} \label{eq:18}
    W_{R_{reduced}} = \prod_{K>0} N_K  e^{-2 (|b_{R}(K)|^2 + |b_{R}(-K)|^2  )}
\end{equation}

To obtain what the inertial observer observes in the inertial frame, we compare the above form in Eq.\eqref{eq:18} with Eq.\eqref{eq:8} for the reduced Wigner distribution in the Rindler frame and read off weight functions $g_i$ (K) and then using Eqs.\eqref{eq:5}-\eqref{eq:8} one can obtain the corresponding $f(k,k')$. The $g_i$ (K) are found to be of the following form:$g_1$ (K) = $g_4$ (K) = 4 $\pi^2 a  \coth{\frac{\pi |K|}{a}} $, $g_2$ (K) = $g_3$ (K) = 0. Using the inverse Fourier transform, the form of the  Wigner distribution in the Minkowski frame is then found to be
\begin{multline}  \label{eq:19}
W_M = N \exp\bigg\{-2 \int _{-\infty} ^ {+\infty} \int _{-\infty} ^ {+\infty}\frac{dk dk'}{4 \pi^2} a_k ^* a_k \bigg[2\pi \delta (k'-k) - \frac{\theta (kk')}{\pi \sqrt{|k||k'|}} [2 \gamma _c + \psi^0(1-i \frac{\ln(k'/k)}{2\pi})+ 
               \\     \psi^0(1+i \frac{\ln(k'/k)}{2\pi}) ] \bigg] \bigg\}
\end{multline}
where $\gamma _c$ is the Euler–Mascheroni constant, and $\psi^0$ represents the polygamma function of order 0. The Eq.\eqref{eq:19} represents a Minkowski state that leads to the Rindler vacuum. Surprisingly, the distribution is independent of the acceleration parameter `a', even though the Eq.\eqref{eq:5}-\eqref{eq:8}, which relates f$(k,k')$ and g$(k,k')$, depends on the parameter `a'. The only input regarding `a' that has gone in this calculation is that `a' must be nonzero. It can be seen from the above expression that the two-point function $f(k,k')$, which gives the expectation value of Minkowski particles, has an infrared divergence. This could be handled by the standard procedure of discretizing by considering the system in a finite box. The above expression in Eq.\eqref{eq:19} appears complicated; however, the distribution simplifies and is more useful in terms of Unruh-Minkowski modes. This can be achieved by choosing $p(K)$ = $q(K)$ = 2 $\coth{\frac{\pi |K|}{a}}$ and $r(K)$ = $s(K)$ = 2 $\cosech{\frac{\pi |K|}{a}}$ in Eq.\eqref{eq:10} which leads to $W_{R_{reduced}}$ in Eq.(\ref{eq:11}) to take the form of the Rindler vacuum. Note that this choice corresponds to one of the many possible ways to choose p,q,r,s such that the coefficients in front of $|b_R(K)|^2$ and $|b_R(- K)|^2$ in Eq.\eqref{eq:11} are unity - which is required for the distribution to be Rindler vacuum. With these chosen expressions for p,q,r,s, one can now write the inertial distribution in terms of Unruh-Minkowski modes using Eq.\eqref{eq:10} to obtain the following form,
\begin{equation} \label{eq:20}
    W_{UM} = \prod_{\xi} N_\xi \exp \Bigg[ -2 (|\alpha _\xi |^2 + |\alpha _{- \xi} |^2 ) \coth{(\xi/2)} - 2 \frac{\alpha_\xi \alpha _{-\xi} + \alpha_\xi ^* \alpha _{-\xi} ^*}{\sinh{(\xi/2)}}  \Bigg] 
\end{equation}
Interestingly, such a choice leads to a form of Wigner distribution, with the corresponding number density of particles turning out to be Planckian, as can be seen by using Eq. \eqref{eq:13},
\begin{equation*}
     \langle \tilde{n}_{\alpha _{\xi}} \rangle =  \langle \tilde{n}_{\alpha _{-\xi}} \rangle = \frac{1 }{e^{\xi}-1}
\end{equation*}
Due to the Planckian form, one may be tempted to associate thermality with the distribution. However, one can check that it cannot be derived from a density operator of the form $e^{- \beta \hat{H}}$ where $\hat{H}$ is the Hamiltonian of the Unruh-Minkowski particles. The precise form of the anomalous averages such as $\langle \alpha _{\xi}^* \alpha _{-\xi}^* \rangle$ breaks this equivalence. For each Rindler frequency, there are two right-moving positive frequencies in Unruh modes. Thus, as mentioned above, there are several other choices of Unruh distributions that lead to the Rindler vacuum. A comparison of the Rindler reduced Wigner distribution in Eq.\eqref{eq:18} with the form in Eq.\eqref{eq:10} provides a general relation between $p(\xi)$, $q(\xi)$, $r(\xi)$ and $s(\xi)$ that leads to the Rindler vacuum, as
\begin{equation} \label{eq:21}
    (p q - r s ) \sinh{(\xi /2)} = p e^{- \xi /2} + q e^{ \xi /2} - r - s
\end{equation}
The choice of r = s = 0 in the above Eq.\eqref{eq:21} \---\ which corresponds to zero anomalous averages \---\  yields the following relationship between p and q:
\begin{equation} \label{eq:22}
    p = \frac{q e^{\xi/2}}{q \sinh{\xi/2} - e^{-\xi/2}}
\end{equation}
By substituting the above relationship, denoted as Eq.\eqref{eq:22}, into Eq.\eqref{eq:14}, one can obtain the expression:
\begin{equation} \label{eq:23}
    \langle \tilde{n}_{\alpha _\xi} \rangle  = - ( \langle \tilde{n}_{\alpha _{-\xi}} \rangle + 1) e^{-\xi}
\end{equation}
Although this relation, denoted as Eq.\eqref{eq:23}, may initially appear to imply the Rindler vacuum to be unphysical, when r=s=0, due to the negative sign, it is indeed physically valid since one only needs the expectation value of the energy density to be bounded from below. One can refer \cite{Ford:2009vz, Ford:2013kga, Emelyanov:2017lbt} for the discussion of the negative energy density in the context of the Rindler observer. We provide two particular examples displayed in Fig.\ref{fig:1}. It can be observed in both examples that \( \langle \tilde{n}_{\alpha _\xi} \rangle \) and \( \langle \tilde{n}_{\alpha _{-\xi}} \rangle \) are bounded from below. Specifically, the minimum of \( \langle \tilde{n}_{\alpha _\xi} \rangle \) and \( \langle \tilde{n}_{\alpha _{-\xi}} \rangle \) in the plot in the left panel of Fig.\ref{fig:1} is \(-1/2\), while the minimum of \( \langle \tilde{n}_{\alpha_{\xi}} \rangle \) is \(-1/4\) and \( \langle \tilde{n}_{\alpha_{-\xi}} \rangle \) remains positive for the right panel. Hence, the corresponding energy density obtained by relation E = (\( \langle \tilde{n} \rangle \) + 1/2 ) $\omega$ is positive. The lower energy density required for the Rindler vacuum, compared to the standard Minkowski vacuum, can be attributed to the relatively weaker constraint implied by quantum inequalities on the world lines of accelerated observers, in contrast to inertial observers, as described in\cite{Ford:2009vz, Ford:2013kga}.

\subsection{A near-Minkowski vacuum functional}
Next, we consider a Minkowski Wigner functional, which represents a state \textit{`close'} to the Minkowski vacuum but not exactly the vacuum. This state can be thought of as a genuine source in the Minkowski frame, such as the ones considered in the laboratory in the case of analogue models, or these may arise as some frequency-dependent noise in the system, which is difficult to remove in an experimental setup. 

To construct one choice of such a distribution, we choose a weight function $f(k,k')$ to be highly peaked in the momentum space and parameterized by a non-negative parameter $\gamma$. The motivation behind choosing such a two-point function $f(k,k')$ is that, in an appropriate limit, the weight function $f(k,k')$ will reduce to a Dirac Delta distribution which leads to the Minkowski vacuum state functional in subsection 3.1. Hence, by introducing a non-negative parameter $\gamma$ such that $f(k,k')$ is highly peaked, one can construct a Minkowski functional, which slightly deviates from the vacuum state as shown below. We choose $f(k,k')$ to be
\begin{equation} \label{eq:24}
    f(k,k') =  \sqrt{\frac{4\pi}{\gamma}} \frac{\theta(kk')}{\sqrt{|k||k'|}} e^{-(\ln(k'/k))^2/\gamma} .
\end{equation}
One can note that this choice corresponds to an approximation of the Dirac Delta distribution and in the limit $\gamma \rightarrow 0$ reduces to the Dirac Delta distribution. The Wigner distribution in the inertial frame is then
\begin{equation} \label{eq:25}
W_M = N \exp\left\{-2 \int _{-\infty} ^ {+\infty} \int _{-\infty} ^ {+\infty}\frac{dk dk'}{4 \pi^2} a_k ^* a_k \sqrt{\frac{4\pi}{\gamma}} \frac{\theta(kk')}{\sqrt{|k||k'|}} e^{-(\ln(k'/k))^2/\gamma} \right\}
\end{equation}
\begin{figure*}[ht!]
        \includegraphics[width=\textwidth]{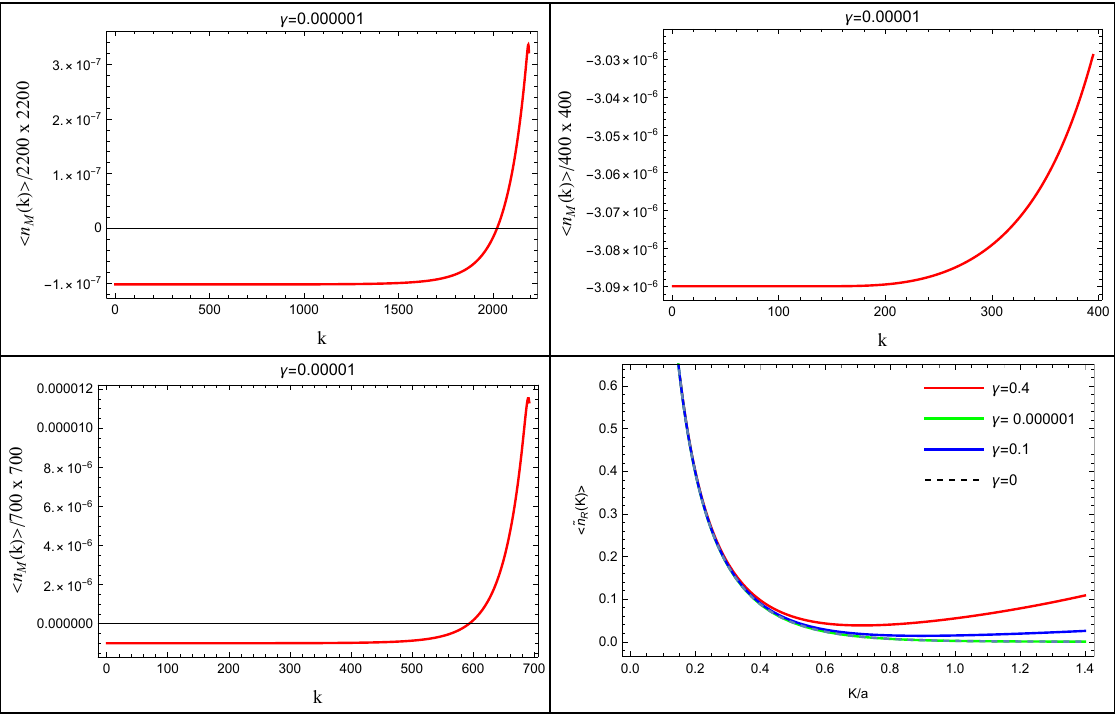}\hfill
        \caption{The first three plots display the expectation value of the number density in the inertial frame, $ \langle n_M(k) \rangle$, while the right bottom panel shows the corresponding $\langle \tilde{n}_R(K) \rangle$ in the Rindler frame}   
        \label{fig:2}
\end{figure*}
\\
As a consistency check, the above expression reduces to the vacuum state for an inertial observer in the limit $\gamma \rightarrow $ 0. To obtain the effect of a small perturbation of $\gamma$ away from 0, we look at the expectation value of number density( $\langle n_M(k) \rangle$). To proceed, one needs to obtain the inverse of $f(k,k')$ to evaluate Eq.\eqref{eq:13}. However, the form chosen for $f(k,k')$ is non-diagonal. Hence, we employ an approximation for the inversion, proceeding with a Taylor series expansion of $f(k, k'; \gamma)$ around $\gamma = 0$. The detailed calculation of $\langle n_M(k) \rangle$ is presented in Appendix B. In the first-order approximation with respect to $\gamma$, the expression for $\langle n_M(k) \rangle $ is given by $-\frac{\gamma \delta (0)}{32} + \mathcal{O} (\gamma ^2)$. Here, $\langle n_M(k) \rangle$ may seem independent of $k$. However, one must note that such a form is contingent not only on the limit $\gamma \rightarrow 0$ but also on the condition imposed on the frequency to be not sufficiently large to render $\gamma k k'$ significant. Here $\delta$(0) is a consequence of the continuum limit of modes. One can divide both sides by the volume of the box to get rid of this divergent term -viz. the expectation value of particles per unit frequency per unit volume is finite. To comprehend the negative sign, one can multiply $\langle n_M(k) \rangle$ by the frequency and then add $\omega /2$ to obtain the energy density in natural units \---\ as explained in the last subsection. In particular, as $\gamma$ $\rightarrow$ 0, the resulting energy density is slightly less than $\omega$/2 but remains positive. 

To investigate the high-frequency behavior, we turn to numerical techniques to find the inverse of $f(k,k')$ using its discrete form with a finite box size in Mathematica. We display the numerical results in Fig\ref{fig:2}. It can be seen from the first three plots in Fig.\ref{fig:2} that the expectation value of number density in an inertial frame is negative at small frequencies\---\ consistent with what we get by the analytical approximation \---\ and diverges at high frequencies for a finite $\gamma$. By choosing smaller values of $\gamma$, we find that the expectation value of number density per unit volume per unit frequency range starts diverging at frequencies higher than for the corresponding case with a comparatively higher $\gamma$. In the strict limit $\gamma \rightarrow 0$, $ \langle n_M(k) \rangle $ approaches zero for all finite frequencies \---\ as expected for the vacuum state, since the divergence pattern as seen in the Fig.\ref{fig:2} will be strictly at infinite value of the frequency. 

To obtain the expectation values of observables for the functional in Eq.(\ref{eq:25}) in the Rindler frame, we compute the reduced Wigner functional for a nonzero acceleration parameter $a$. By substituting the two-point function $f(k,k')$ from Eq. \eqref{eq:22} into Eq. \eqref{eq:4} to \eqref{eq:7}, one obtain the following $g_s$: $g_1(K) = g_4(K) = 4 \pi ^2 a \exp{(-\gamma K^2/4 a^2)}$, and $g_2(K) = g_3(K) = 0$. These lead to the reduced Wigner functional in \eqref{eq:8} to take the following form
\begin{equation} \label{eq:26}
    W_{R_{reduced}} = \prod_{K>0} N_K \exp \left\{-2 (|b_{R}(K)|^2 + |b_{R}(-K)|^2  ) e^{-\gamma K^2/4a^2} \tanh{\frac{\pi K}{a}} \right\}
\end{equation}
Using Eq.\eqref{eq:12} we get the following expectation value of the number density in the Rindler frame:

\begin{equation} \label{eq:27}
    \langle \tilde{n}_R(K) \rangle = \frac{ 1}{2} \bigg( e^{\frac{\gamma K^2}{4a^2}} \coth{\frac{\pi K}{a}} -1 \bigg).
\end{equation}
As expected, in the limit $\gamma \rightarrow 0$, the reduced Rindler Wigner functional turns out to be thermal, corresponding to the Minkowski vacuum. But for a nonzero $\gamma$, the expectation value of the number density (see Eq. \eqref{eq:12}) diverges at high frequencies for ( see the right panel of Fig.\ref{fig:2}) $\gamma$ comparable or greater than $a/K$. This divergence can be regulated by choosing a frequency cutoff. However, the corresponding number density distribution may or may not be regulated by the same cutoff in the inertial frame as well as in a different frame. The form of divergence may depend on the state as well as the frame of reference, and further, a small perturbation around a given state can change the regularization significantly in different frames. \\

In the same context, one can analyze the large frequency behavior of the number density expectation value in terms of the Unruh Minkowski modes. To re-express the Minkowski functional in Eq.\eqref{eq:25} in terms of the UM mode functional in an inertial frame we compare Eq.\eqref{eq:10} with Eq.\eqref{eq:24} which suggests identifying $p(\xi) = q(\xi)$ = 2 $e^{-\gamma \xi ^2 /16 \pi ^2}$ under a choice r = s = 0. Thus, the weight functions in the Wigner functional in terms of the Unruh Minkowski degrees of freedom also have a Gaussian form. Using Eq.\eqref{eq:12}, the expectation value of the number density of Unruh-Minkowski particles for smaller frequencies, to the linear order in $\gamma$, is found to be $\gamma \xi ^2 / 32 \pi ^2 + \mathcal{O}(\gamma ^2)$. However, for larger frequencies, it grows rapidly. One concludes from here that in terms of Unruh particles, the $\langle \tilde{n}_{\alpha_\xi} \rangle$ and $\langle \tilde{n}_{\alpha_{-\xi}} \rangle$ both diverge at high frequencies for a $\gamma$ greater than $32 \pi^2/\xi ^2$.
The functional form of the divergence associated with number density is sensitive to the form of $f(k,k')$ chosen. Even with a choice of $f(k,k')$, which slightly deviates from the corresponding form of the vacuum, the $\langle \tilde{n}(k) \rangle $ may be affected in a nontrivial way. For example, let us consider the following form
 \begin{equation} \label{eq:28}
     f(k,k')=\frac{\pi}{c} \frac{\theta(kk')}{\sqrt{|k||k'|}} e^{-|c \ln(k'/k))|} 
 \end{equation}
 for some nonzero constant `c'. The corresponding Rindler frame occupation number 
\(\langle \tilde{n}_R(K) \rangle\), is found to be such that its second derivative becomes constant at high frequencies and goes to \(0\) at lower frequencies.

\subsection{A peaked frequency distribution}

\begin{figure*}[ht!]
        \includegraphics[width=\textwidth]{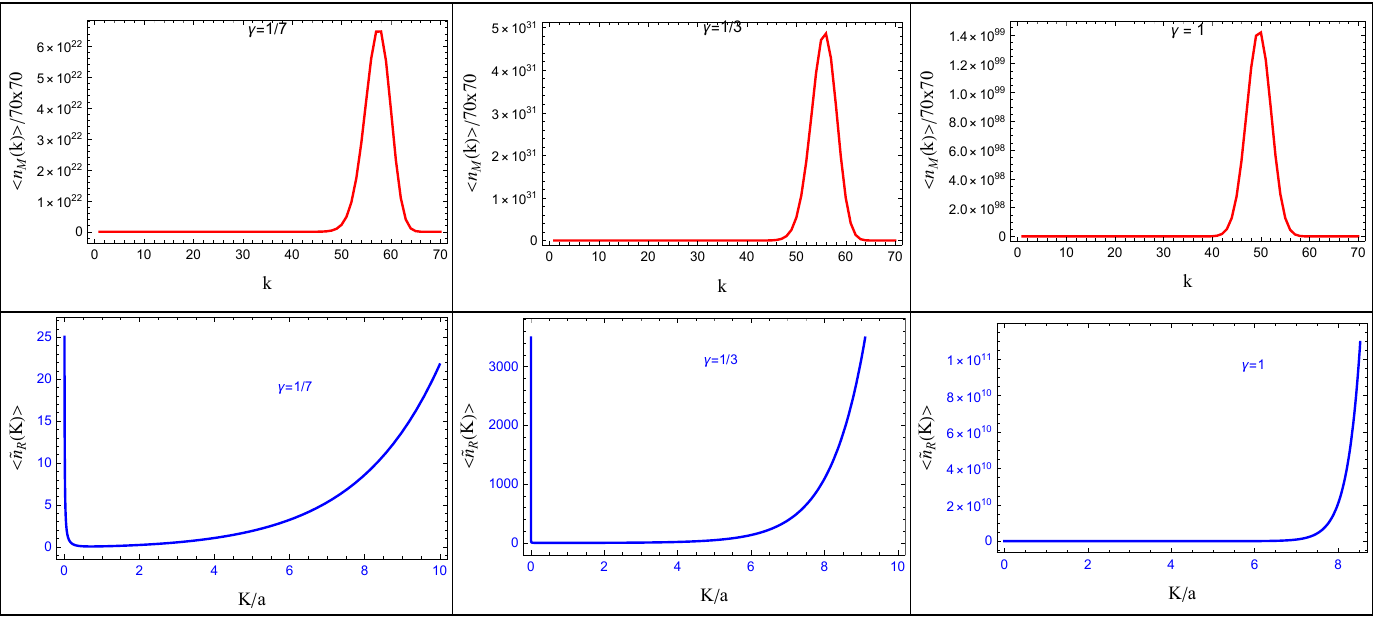}\hfill
        \caption{The red curve plots display the expectation value of the number density in the inertial frame $\langle \tilde{n}_M(k) \rangle$ for the state given in section 3.4 while the blue curves show the corresponding $\langle \tilde{n}_R(K) \rangle$  in the Rindler frame.}    
        \label{fig:3}
\end{figure*}
In the earlier subsections, we discussed the vacuum state Wigner functionals with its weight function, where $f(k,k')$ is sharply peaked in the Dirac delta distribution sense about the origin in momentum space. We also analyzed a case of Wigner functional, which slightly deviates from the vacuum state functional by considering its corresponding weight function $f(k,k')$ to have a small width around a peak value in momentum space. Here, we consider a particular Wigner functional such that the expectation value of the corresponding Minkowski number density in momentum space is sharply peaked instead of the sharply peaked two-point weight function $f(k,k')$. One can note that Eq.(\ref{eq:13}) hints that the expectation value of number density and $f(k,k')$ are inversely related, and hence the peak in each separately has a different interpretation. One can interpret a peak in the number density of a state as corresponding to a band pass filter, which allows only particles having frequencies lying about some peak value to pass through. Furthermore, it can also be used as a toy model for Bose-Einstein condensates, high-density gluon condensates such as color glass condensate, pion condensates, or for information processing in quantum computing, where a distribution of particles with nearly the same frequency is considered.  

To construct such a state we choose the following form for the inertial frame Wigner distribution:
\begin{equation} \label{eq:29}
W_M = N \exp(-2 \int _{-\infty} ^ {+\infty} \int _{-\infty} ^ {+\infty}\frac{dk dk'}{4 \pi^2} a_k ^* a_k \frac{2 \theta(kk')}{\gamma \sqrt{kk'} \left( (k/k')^{1/2\gamma} + (k'/k)^{1/2\gamma}  \right)} ) ,
\end{equation}
where $\gamma $ is a non-negative parameter. Such a choice leads the expectation value of the Minkowski number density operator $\langle \tilde{n}_M(k) \rangle$ to be peaked in the momentum space as described below. We follow the procedure outlined in the previous subsection to numerically obtain the plots of inertial frame number density $\langle \tilde{n}_M(k) \rangle$ and Rindler frame number density  $\langle \tilde{n}_R(K) \rangle$ as shown in Fig.\ref{fig:3}. The peak of number density $\langle \tilde{n}_M(k)$ in frequency space shifts with a change in the value of the parameter $\gamma$. From the plots, one can observe that the position of the peak shifts rightward, that is, towards a higher frequency with a decrease in $\gamma$. Further, the absolute value of the peak of $ \langle \tilde{n}_M \rangle$ also decreases with a decrease of   $\gamma$. Thus, as $\gamma \rightarrow$ 0, the peak of the number density appears to be shifted to infinity; however, the absolute value of the peak also approaches zero. This behavior is consistent with the expected limit wherein the  $\langle \tilde{n}_M(k)$ approaches the vacuum state  $\langle \tilde{n}_M \rangle$, when $\gamma \rightarrow 0$. Using Eqs. \eqref{eq:5}-\eqref{eq:9}, one obtains the form of $g_i(K)s$ to be $g_1(K) = g_4(K) = 4 \pi^2 a \sech{(\pi \gamma |K|/a)}$ and $g_2(K) = g_3(K) = 0$. The corresponding Rindler frame reduced Wigner distribution for a nonzero `a' is found to be
\begin{equation} \label{eq:30}
    W_{R_{reduced}} = \prod_{K>0} N_K \exp \left\{-2 (|b_{R}(K)|^2 + |b_{R}(-K)|^2  ) \sech{\frac{\pi |K| \gamma}{a}} \tanh{\frac{\pi |K|}{a}} \right\}.
\end{equation}
The corresponding expectation value of the number density in the Rindler frame is obtained to be

\begin{equation} \label{eq:31}
    \langle \tilde{n}_R(K)\rangle = \frac{1}{2} \bigg( \cosh(\frac{\pi \gamma |K|}{a}) \coth(\frac{\pi |K|}{a}) -1 \bigg).
\end{equation}
For $\gamma = 0$, $\langle \tilde{n}_R(K) \rangle$ reduces to the standard Planckian form with the Unruh temperature T = a/$2\pi$, as is expected since $\gamma \rightarrow 0$ represents the vacuum Minkowski-particle state.
The bottom panel in Fig.\ref{fig:3} shows the expectation value of Rindler number density $\langle \tilde{n}_R(K) \rangle$ given in, Eq.\eqref{eq:31}. One can see that it diverges at higher frequencies. The divergence becomes sharper as $\gamma$ increases from 1/7 to 1. Whereas at low-frequencies, the behavior of  $\langle \tilde{n}_R(K) \rangle$ is finite and seems approaching zero for the case of $\gamma $ = 1, while for $\gamma $ = 1/3 and 1/7 there is a rise in  $\langle \tilde{n}_R(K) \rangle$ when the Rindler frequency approaches zero. These observations suggest that an increase in $\langle \tilde{n}_R(K)\rangle$ for the field in this state when $\gamma \rightarrow < 1/7$ starts slightly earlier than the same for $\gamma \rightarrow > 1$. Alternatively, one can conclude that as $\gamma \rightarrow 0$, $\langle \tilde{n}_R(K) \rangle$ diverges at lower frequencies and decays exponentially at high frequencies, while the behavior is opposite for a relatively high $\gamma$, if one restricts oneself to a finite range of frequencies.

\subsection{Indistinguishability of thermal and quantum fluctuations }
Previous works in the literature have attempted to explore the relationship between quantum and gravitational phenomenon by exploring the relation or connection between quantum and statistical fluctuations \cite{Smolin_1986}. There are also proposals to describe inertia as a consequence of the second law of thermodynamics \cite{inertia}. In \cite{Kolekar:2013aka}, the indistinguishability of thermal and quantum fluctuations was demonstrated in the context of the Unruh thermal bath. Using the density matrix formalism, the corresponding reduced density matrix in the Rindler frame was obtained starting from a thermal bath of one set of Unruh Minkowski particles. The resultant $\rho_{red}$ in \cite{Kolekar:2013aka, Bruschi:2013tza} was symmetric in the temperature $T_{UM} $ of the thermal bath considered and the Unruh acceleration temperature T= a/2 $\pi$. Here, we explore further this relationship in the Wigner functional formalism. In \cite{Kolekar:2013aka}, $\langle \tilde{n}_{\alpha_{-\xi}} \rangle$ is taken to be zero, that is, thermality is considered only for $\langle \tilde{n}_{\alpha_{\xi}} \rangle$. We consider a Wigner functional that represents a thermal bath of Unruh-Minkowski particles, where both $\langle \tilde{n}_{\alpha_\xi} \rangle$ and $\langle \tilde{n}_{\alpha_{-\xi}} \rangle$ follow a thermal distribution.  

We consider the following Unruh Minkowski Wigner distribution:

\begin{equation} \label{eq:32}
    W_{UM} = \prod_{\xi} N_\xi e^{-2 |\alpha_{\xi}|^2 \tanh{\frac{a |\xi| }{2\gamma}} - 2 |\alpha_{-\xi}|^2 \tanh{ \frac{a |\xi|}{2\gamma}}} = \prod_{K>0} N_K e^{-2 |\alpha_{K}|^2 \tanh{\frac{\pi K}{\gamma}} - 2 |\alpha_{-K}|^2 \tanh{ \frac{\pi K}{\gamma}}} ,
\end{equation}
where $\gamma$ is some nonnegative real parameter. It is known that different quantum states can lead to the same expectation value of, say, the number density operator, which is a two-point correlator, while the corresponding expectation values of higher order correlations of the same operator, which are essentially fluctuations, can be different. In subsection 3.2, we found a  Planckian expectation value of Unruh Minkowski particles $\langle n_{UM}(\xi)\rangle$ for the state functional in Eq.\eqref{eq:20} that leads to the corresponding reduced Wigner functional in the Rindler frame to be that of the Rindler vacuum. In contrast, the Wigner functional in Eq.\eqref{eq:32}  represents a true thermal bath, with both $\langle n_{\alpha_\xi} \rangle$ and $\langle n_{\alpha_{-\xi}} \rangle$ describing a thermal bath at the same temperature, characterized by $\gamma$. By substituting $p'(K)=q'(K)= \tanh{\pi |K|/\gamma}$ and $r = s= 0$ in equation \eqref{eq:11}, we obtain the following reduced Wigner distribution in the Rindler frame for a nonzero acceleration parameter $a$:
\begin{equation} \label{eq:33}
W_{R_{\text{reduced}}} = \prod_{K>0} N_K \exp \left\{-2 (|b_{R}(K)|^2 + |b_{R}(-K)|^2 ) \tanh{\frac{\pi K}{\gamma}}\tanh{\frac{\pi }{a}} \right\}
\end{equation}
which, using  Eq.\eqref{eq:12}, yields the following expectation value of number density in the Rindler frame:
\begin{equation} \label{eq:34}
\langle \tilde{n}_R(K) \rangle = \frac{1 }{(e^{2\pi K/\gamma}-1)(1-e^{-2\pi K/a)}} + \frac{1}{(e^{2\pi /a}- 1)(1-e^{-2\pi K/\gamma})}.
\end{equation}
We first note that the above expression is different from the expectation value $\langle \tilde{n}_R (K) \rangle$ for the single thermal bath of $\langle \tilde{n}_{\alpha_\xi}\rangle$ considered in \cite{Kolekar:2013aka}. However, the expression above, Eq. \eqref{eq:34}, is the same as the expression for the expectation value of the number density given in Eqs. (28) and (40) of \cite{rindrind}, with the identification $\beta = 2 \pi / \gamma$ and `g' = `a'. In \cite{rindrind}, along with the Bogoliuobov calculations, the authors have also illustrated the response rate of a Unruh-DeWitt detector for a Rindler Rindler trajectory in the Minkowski vacuum. One can interpret the result in Eq.\eqref{eq:34} as the sum of expectation values of the number density of two gray bodies with the gray body factor equal to the partition function of the other one. One can also note that Eq.\eqref{eq:33} is invariant under the exchange of two temperatures $T_1 = a/2\pi $ and $ T_2 = \gamma /2\pi $. In other words, the reduced Wigner functional with acceleration parameter `a' for a thermal bath at temperature $\gamma /2\pi$ is identical to the reduced Wigner functional with acceleration parameter corresponding to Unruh-Davies temperature $\gamma /2\pi$ for a thermal bath of UM particles at temperature a/$2\pi$. This property supports the idea of  indistinguishability of thermal and quantum fluctuations \cite{Kolekar:2013aka}, which tells that within the domain of thermodynamic experiments, the quantum fluctuations and statistical fluctuations are indistinguishable. 

We now proceed further to obtain an inertial frame functional in terms of Minkowski modes that yields the reduced Wigner functional Eq.\eqref{eq:33}. We first read off  $g_1(K) = g_4(K) = 4 \pi^2 a \tanh{(\pi |K|/\gamma)}$ and $ g_2(K)=g_3(K) = 0 $  by comparing Eq.\eqref{eq:33} with Eq.\eqref{eq:9}. Next, we use the inverse Fourier transform in Eq.(\ref{eq:5} - \ref{eq:8}) to determine the weight function f$(k,k')$.  Substituting the resulting weight function into Eq.\eqref{eq:3} yields
\begin{multline} \label{eq:35}
     W_M = N \exp\bigg[-2 \int _{-\infty} ^ {+\infty} \int _{-\infty} ^ {+\infty}\frac{dk dk'}{4 \pi^2} a_k ^* a_k [2 \pi \delta(k-k') + \frac{\gamma' \theta (kk')}{2 \pi \sqrt{|k||k'|}} [-\psi^0(1-i \frac{\gamma' \ln(k'/k)}{4\pi})  \\  -\psi^0(1+i \frac{\gamma' \ln(k'/k)}{4\pi}) + \psi^0(\frac{1}{2}-i \frac{\gamma' \ln(k'/k)}{4\pi})+ \psi^0(\frac{1}{2}+ i \frac{\gamma' \ln(k'/k)}{4\pi})] ]\bigg]
\end{multline}
where, $\gamma' =\gamma/a$. Here, $\psi^0$ represents the polygamma function of order 0, i.e., the digamma function. The range of the integration over k and $k'$ in the above expression is - $\infty$ to $\infty$. However, one can split the integration and use the $\theta (kk')$ term to change the integration limit from 0 to $\infty$. This yields some another weight, say h, where the integration runs from 0 to $\infty$. The expectation value of the number density of Minkowski particles can be determined using Eq.\eqref{eq:12}. To obtain the inverse of $ h(k,k')$ we use the definition $\int h(k,k')h^{-1} (k',k'') dk' $ = $\delta(k-k'')$ and substitute the form of $h(k,k')$ from Eq.\eqref{eq:35} to identify the resultant expression as a Fredholm integral equation of the second kind \cite{lovitt2014linear}. Using the standard solution of the Fredholm integral equation in Eq.\eqref{eq:13} yields the following series---describing the expectation value of the number density of Minkowski particles---which is convergent as $\gamma' \rightarrow 0$ .
\begin{multline} \label{eq:36}
    \langle n_M(k) \rangle = \frac{1 }{2} \bigg[ - \frac{\gamma'}{2\pi} h(k,k) + \frac{\gamma'^2} {4\pi ^2} \int _{0^+} ^\infty h(k,k') h(k',k) dk' - \\ \frac{\gamma'^3} {8\pi ^3} \int _{0^+} ^\infty h(k,k') \int _{0^+} ^\infty h(k',k'')h(k'',k) dk' dk'' +.....\bigg]
\end{multline}
where $h(k,k')$ represents the term, apart from $\theta (kk')$, in the weight function $f(k,k')$ that appears alongside the Dirac delta function, given by \\ 
\begin{multline} \label{eq:37}
h(k,k') = \frac{1}{2 \pi \sqrt{|k||k'|}}\bigg[-\psi^0(1-i \frac{\gamma' \ln(k'/k)}{4\pi}) -\psi^0(1+i \frac{\gamma' \ln(k'/k)}{4\pi}) + \psi^0(\frac{1}{2}-i \frac{\gamma' \ln(k'/k)}{4\pi})+ \\ \psi^0(\frac{1}{2}+ i \frac{\gamma' \ln(k'/k)}{4\pi})\bigg]  .
\end{multline}
Additionally, one can confine the system to a finite box, which will ensure the lower limit of integration to be $0^+$. For $\gamma'$ = 0 the $\langle n_M(k) \rangle$ in Eq. \eqref{eq:37}, yields the vacuum state in Minkowski and the $\langle n_R(K) \rangle$ in Eq.\eqref{eq:34} reduces to Planckian distribution in the Rindler frame, as expected. Another interesting aspect would be to explore the effect of the extra term $h(k,k')$ to first order in $\gamma'$. Using $\lim_{x \to 0} [-\psi^0(1-i \frac{\gamma' x}{4\pi}) -\psi^0(1+i \frac{\gamma' x}{4\pi}) + \psi^0(\frac{1}{2}-i \frac{\gamma' x}{4\pi})+ \psi^0(\frac{1}{2}+ i \frac{\gamma' x}{4\pi})] $ = -2.77259 and retaining terms linear in $\gamma'$ in Eq.\eqref{eq:36} we obtain the following expression for the number density of Minkowski particles: 
\begin{equation} \label{eqn:37}
    \langle n_M(k) \rangle = 0.035115 \frac{\gamma'}{|k|} .
\end{equation}
Interestingly, for small values of $|k|$, the above expression of $ \langle n_M(k) \rangle $ exhibits a dependence on |k| akin to that of a bath with Planckian distribution. 

Further, it is worth mentioning that in the case when we choose $p'(K) = \tanh(\beta |K| / 2)$ and $q'(K) = 1$ in Eq.\eqref{eq:10}, the corresponding configuration represents a thermal bath of Unruh-Minkowski (UM) particles in $\langle \tilde{n}_{\alpha _{\xi}}\rangle$, and with a zero average particle number density for $\langle \tilde{n}_{\alpha _{-\xi}} \rangle$, which is the specific scenario discussed in reference \cite{Kolekar:2013aka}. Remarkably, this case is also consistent with detector response rate calculations \cite{Barman:2021oum}. However, it's crucial to recognize that, since a particle detector also gets extra contributions apart from particle-like excitations, in more general cases, detector response may lead to varying conclusions, as demonstrated in different scenarios in reference \cite{Sriramkumar:1999nw}.

\subsection{Fermionization of a bosonic field}
In curved spacetime, a relationship between statistics and spin naturally arises from the underlying spacetime dynamics— a connection that is absent for inertial observers in the flat geometry of Minkowski spacetime. \cite{Verch:2001bv, parker, Good:2012cp}. However, the nontrivial Bogoliubov coefficients for an accelerated observer lead to the emergence of the spin-statistics connection for accelerated observers even in Minkowski spacetime \cite{Good:2012cp}. Additionally, the equivalence principle, which posits a local equivalence between acceleration and gravitational effects, also suggests that this connection persists in Rindler space. The response rate calculation in \cite{10.1143/PTP.88.1} for a detector moving on a uniformly accelerated trajectory shows that the statistics of the Minkowski vacuum of a massless scalar field are Fermionic in an odd number of dimensions and are bosonic for even number of dimensions. However, the Bogoliubov calculation yields a Bosonic distribution in any dimension for the number density expectation in terms of the uniformly accelerated observer operators in a Minkowski vacuum \cite{Padmanabhan:2019art}. In this subsection, we consider a Minkowski Wigner functional which curiously leads to a mixture of both Fermionic and bosonic distributions in the Rindler frame, as described below. 

We define a Minkowski state by modifying the weight function $f(k,k')$ discussed in the preceding subsection in Eq.\eqref{eq:35} and introducing a change in sign of all the terms other than the Dirac delta in the weight function $f(k,k')$. The objective is to assess the influence of the corresponding remaining term, h$(k,k')$. More precisely, we take the Wigner distribution for the inertial frame to be of the following form:
\begin{multline} \label{eq:38}
     W_M = N \exp \bigg[-2 \int _{-\infty} ^ {+\infty} \int _{-\infty} ^ {+\infty}\frac{dk dk'}{4 \pi^2} a_k ^* a_k [2 \pi \delta(k-k') - \frac{\gamma' \theta (kk')}{2 \pi \sqrt{|k||k'|}} [-\psi^0(1-i \frac{\gamma' \ln(k'/k)}{4\pi}) \\  -\psi^0(1+i \frac{\gamma' \ln(k'/k)}{4\pi}) + \psi^0(\frac{1}{2}-i \frac{\gamma' \ln(k'/k)}{4\pi})+ \psi^0(\frac{1}{2}+ i \frac{\gamma' \ln(k'/k)}{4\pi})] ] \bigg] .
\end{multline}
\\
Here, $\psi^0$ represents the polygamma function of order 0, i.e., the digamma function. The expectation value of the number density in the inertial frame can be obtained by the similar procedure as in the preceding subsections. It yields a series that differs from Eq.\eqref{eq:36} just by a negative sign, and it is convergent for $\gamma \rightarrow 0 $, if one considers the system to be in a box. However, in the present scenario, the expectation value of Minkowski particles up to the first order in $\gamma$ turns out to be $ \langle n_M(k) \rangle = - 0.035115 \gamma/|k|$. We note that this value is negative of the $\langle n_M(k) \rangle$ in Eq.\eqref{eqn:37}. Once again, we can multiply $\langle n_M(k) \rangle$ by frequency and add the zero point energy to get the positive and finite answer. One can get the weight function corresponding to the inertial frame Wigner functional displayed in Eq.\eqref{eq:38} by substituting $g_1(K) = g_4(K) = 4 \pi^2 a (2-\tanh{(\pi |K|/\gamma)}), g_2(K)=g_3(K) = 0 $ in Eq.\eqref{eq:4}-\eqref{eq:8} and using the inverse Fourier transform. The substitution of these $g_s$ in Eq.\eqref{eq:9} yields the following reduced Wigner distribution in the Rindler frame for a nonzero `a':\\

\begin{equation} \label{eq:39}
    W_{R_{reduced}} = \prod_{K>0} N_K \exp \left\{-2 (|b_{R}(K)|^2 + |b_{R}(-K)|^2  ) (2-\tanh{\frac{\pi |K|}{\gamma }}) \tanh{\frac{\pi |K|}{a}} \right\}.
\end{equation}
Using \eqref{eq:12} for the expectation value of number density one obtain
\begin{eqnarray} \label{eq:40}
    \langle \tilde{n}_R(K) \rangle = \frac{1 }{-1 + e^{2 \pi |K|/a}}  - \frac{1}{3 + e^{2 \pi |K|/\gamma }} - \frac{2 }{(-1 + e^{\frac{2 \pi |K|}{a}})(3 + e^{\frac{2 \pi |K|}{\gamma}})},
\end{eqnarray}
where $\gamma =\gamma' a$. In the above expression, the second term exhibits the Fermionic statistics. The overall structure of the expression, with the exception of the Fermionic form instead of bosonic component is similar to, Eq.(9) of \cite{Kolekar:2013aka}, where the authors found the expectation value of the number density in the Rindler frame corresponding to a Unruh thermal bath in the Minkowski frame, to be the sum of two bosonic distributions plus an additional term that is the product of the first two terms. One can interpret Eq.\eqref{eq:40} in terms of the spontaneous and stimulated emission of Rindler particles by writing as $\langle \tilde{n}^\beta \rangle $ + $\langle \tilde{n}^{\beta '} \rangle $ + 2 $\langle  \tilde{n}^\beta \rangle $ $\langle \tilde{n}^{\beta '} \rangle $. It's important to note that the Fermionic nature of Rindler noise was previously known in the case of odd dimensions for the Unruh Dewitt detector response \cite{10.1143/PTP.88.1}. Here, a Fermionic component arises in even dimensions in spite of using the Bogoliuobov calculation, which goes into calculating the Rindler reduced Wigner functional from the Minkowski functional.

As mentioned earlier, one can obtain the reduced Wigner functional described in Eq.\eqref{eq:39} by using several other inertial frame Unruh Minkowski distributions. We arrive at one such interesting inertial frame Wigner functional, in terms of Unruh-Minkowski modes, that yields Eq.\eqref{eq:40} by substituting the following weight functions in Eq. \eqref{eq:10}:
\begin{align*} \numberthis \label{eq:41}
    p(\xi) = & q(\xi) = \sech{\left(\frac{\xi}{2}\right)}, \\
    r\left( \xi \right) = & \frac{2 e^{\frac{\xi}{2 \gamma'}} j(\xi)}{3 + e^{\frac{\xi}{\gamma'}}}, \\
    s\left( \xi \right) = &  -\frac{e^{-\frac{\xi}{2 \gamma'}} \, \text{sech}\left(\frac{\xi}{2}\right) \left(2 + e^{\frac{\xi}{\gamma'}} \left(-2 + \text{sech}\left(\frac{\xi}{2}\right)\right) + \text{sech}\left(\frac{\xi}{2}\right)\right) \left(-2 + \tanh\left(\frac{\xi}{2 \gamma'}\right)\right)}{2 j(\xi)}, \\
    j(\xi) = &   2 e^{- \xi /2 \gamma'} + \frac{\cosh\left(\frac{\xi}{2 \gamma'}\right)}{\left(1 + e^{\frac{\xi}{\gamma'}}\right)^{3/2}} \left\{
    \left( 3 + 4 e^{\frac{\xi}{\gamma'}} + e^{\frac{2 \xi}{\gamma'}} + 8 \cosh\left(\frac{\xi}{2}\right) \right) \right.  \times \\ 
    &  \left. \left( -3 + 2 \cosh\left(\frac{\xi}{2}\right) + e^{\frac{\xi}{\gamma'}} \left( -1 + 2 \cosh\left(\frac{\xi}{2}\right) \right) \right) \text{sech}^2\left(\frac{\xi}{2}\right) \right\}^{1/2} ,
\end{align*}
\\
where $\gamma' = \gamma/a$. Substituting the above weight functions, denoted as Eq.\eqref{eq:41}, in Eq.\eqref{eq:14} we obtain the expectation value of the Unruh Minkowski number density as
\begin{eqnarray} \label{eq:42}
    \langle \tilde{n}_{\alpha_{\xi}} \rangle = \langle \tilde{n}_{\alpha_{- \xi}} \rangle = \frac{1}{e^{2 \pi |K| / \gamma} - 1},
\end{eqnarray}
which is nothing but a Planckian distribution. The Bosonic statistics of inertial frame expectation value Eq.\eqref{eq:42} suggests that there can be a mixing of statistics by just choosing a different frame. However, as discussed in subsection 3.2, it does not represent a true thermal bath because of the nonzero r$(\xi)$ and s$(\xi)$ that yield nonzero anomalous averages.

\section{Discussion} \label{discussion}

Using the Wigner distribution formalism, we derived a general form of the reduced Wigner functionals in Rindler spacetime by tracing out the degrees of freedom beyond the Rindler horizon of a real massless scalar field in (1+1) dimensions for a large set of states of Minkowski and Unruh-Minkowski modes. The obtained Wigner functionals can be used to compute the expectation value of any relevant physical observables. We further presented a general expression for the expectation value of the number density operator for each of these Wigner functionals. As a consistency check, we examined the Minkowski vacuum, which yields the reduced Wigner functional of a thermal bath in the Rindler frame \---\ illustrating the standard Unruh effect. Using the general result, we analyzed several special Minkowski distributions and their corresponding reduced Rindler distributions. Below, we highlight some of the significant observations in these cases.  

An interesting Minkowski functional to investigate is a genuine source in the Minkowski frame, such as the ones considered in the laboratory in the case of analogue models, or as a frequency-dependent noise in the system. To explore, we considered a near-Minkowski vacuum functional, with a slight deviation from the Minkowski vacuum characterized by a non-negative parameter \(\gamma\). The expectation value of the Minkowski number density, \(\langle \tilde{n}_M(k)\rangle\), remains constant at lower frequencies but increases at higher frequencies. In the corresponding Rindler space, the number density \(\langle \tilde{n}_R(K)\rangle\) deviates from thermality at frequencies higher compared to the acceleration scale for a finite \(\gamma\). As a result, the occupation number in the Rindler frame matches that of a Planckian bath up to a frequency, whose scale is smaller than the acceleration scale. An observer sensitive only to lower frequencies will still observe thermal behavior for sufficiently small $\gamma$.

We further analyzed a Wigner functional distribution where \(\langle \tilde{n}_M(k)\rangle\) is sharply peaked around a specific Minkowski mode frequency. Such a distribution acts as a toy model for situations such as a band-pass filter, which allows only particles with frequencies near a peak value to pass through, even for analogue systems where the occupation number is peaked around certain energy state like in the case of Bose-Einstein condensates, high-density gluon condensates (such as color glass condensate), pion condensates, or information processing in quantum computing, where a distribution of particles with nearly the same frequency is considered. The peak can be shifted within the \(k\) space by varying the parameter \(\gamma\); however, moving the peak to higher frequencies reduces the overall magnitude of \(\langle \tilde{n}_M(k)\rangle\). For large $\gamma$, we observe a complementary relation between the inertial and Rindler frame such that if we have an inertial frame state with a large occupation number of particles in the small frequency modes, then the Rindler frame observer, with a detector having sensitivity to detect a finite range of frequencies, will find a deviation from the Planckian and will detect most of the particles in the high-frequency range. However, if there are only a few particles at higher frequencies and none at lower frequencies in the inertial frame, the Rindler observer detects a Planckian spectrum in the low-frequency range with small corrections in the high-frequency limit.
 
Another interesting distribution which can be setup in a laboratory for analogue models is a bath of Unruh-Minkowski particles in the inertial frame, such that it exhibits thermal characteristics in both of the Unruh-Minkowski modes that correspond to the same Rindler frequency i.e, the thermality in both \( \alpha_\xi \) and \( \alpha_{-\xi} \). The corresponding reduced Wigner functional is found to be invariant under the exchange between the Unruh-Davies temperature ($T_1 = a/2\pi$)  and the bath temperature in the inertial frame ($T_2 = \gamma /2\pi$). Therefore, an observer, by conducting thermodynamical experiments alone, cannot distinguish whether he is accelerating with an acceleration \( a\) in a thermal bath at temperature \(\gamma/2\pi \) or with an acceleration of \(\gamma\) in a thermal bath at temperature \(a / 2\pi\). Thus, we obtained the indistinguishability of quantum and statistical fluctuations even with a bath of Unruh-Minkowski particles exhibiting thermality in both of the Unruh-Minkowski modes that correspond to the same Rindler frequency, as opposed to the same with only one mode as described in \cite{Kolekar:2013aka}. Remarkably, we pointed out that the occupation number     \(\langle \tilde{n}_R(K)\rangle\), observed by the Rindler observer in this thermal bath, is the same as the occupation number observed by a Rindler-Rindler observer in Minkowski vacuum \cite{rindrind}. 

In \cite{Kolekar:2013aka}, the occupation number of Rindler particles observed by a uniformly accelerated observer in a bath of Unruh-Minkowski particles, which is thermal only in $\alpha_\xi$ for an inertial observer, was shown to be the sum of two Planckian factors with Bosonic statistics, along with a third term that is the product of the first two terms. The authors of \cite{Kolekar:2013aka} interpreted it as the presence of a thermal bath of Unruh Minkowski particles in an inertial frame stimulating an additional emission of Rindler particles following the same bosonic statistics. To further test the connection of simulated emmission and the statistical nature of the distribution of emitted particles, we considered an inertial frame state distribution which yields the occupation number of Rindler particles for an accelerated observer to be the sum of a Planckian factor with Bosonic statistics, a factor with Fermionic statistics, and a third term, which is twice the product of the first two terms. Thus it is possible to add particles in the inertial frame over the Minkowski vacuum such that it stimulates both bosonic and fermionic distribution in terms of Rindler particles in the accelerated frame. One can note that this is still a $1+1$ dimensional example in contrast to \cite{10.1143/PTP.88.1}, where it was shown that depending on the dimensions, one obtains bosonic or fermionic distributions for even and odd dimensions, respectively. 

The special distributions discussed above illustrate the applicability of the general expressions for a set of distributions introduced in the main text and that they can be used as a toy model in the understanding of various physical situations, such as those in the study of the Unruh effect through analog models in a laboratory. Indeed, there are several other distributions where the formalism presented can be applied. Here, we have restricted our analysis to real massless scalar fields in Minkowski space with (1 + 1) dimensions for a special set of states. Extending to other dimensions with different fields in different states may yield further insight that can be explored in future work.

\section*{Acknowledgements}
The authors thank Jorma Louko for his valuable discussions and insightful comments on this manuscript.

\begin{appendices}
\section{\bf{ Wigner Distribution for a bath of Minkowski particles}}\label{appendix A} 
In this Appendix, we briefly outline the calculation of the reduced Wigner distribution for Minkowski modes discussed in Section 2. The real massless scalar field in (1+1) dimensional Minkowski spacetime can be described by the following  general solution of the Klein-Gordon equation
\begin{equation} \label{eq:43}
\hat{\phi}(t,x) = \int _{-\infty } ^{+\infty} \frac{dk}{\sqrt{2 \pi |k|}} [e^{-i|k|ct + ikx} \hat{a}(k) + e^{i|k|ct - ikx} \hat{a}^\dag(k)].
\end{equation}
\\
The above expression suggests that the field can be thought of as an infinite collection of harmonic oscillators, one for each k. One can have a bath filled with such fields completely described by its density matrix or its Wigner distribution. We consider those sets of distributions in an inertial frame that can be described by the Wigner distribution given in Eq. $\eqref{eq:3}$. We also assume Eqs. $\eqref{eq:5}$ - $\eqref{eq:8}$ to be satisfied. Now, using the standard Bogoliubov transformation of Minkowski space creation and annihilation operators from inertial to the Rindler frame \cite{parker}, one gets the following Wigner distribution in the Rindler frame:
\begin{equation} \label{eq:44}
W_R = \bar{N} \exp(-\frac{1}{2\pi ^2} \int _{-\infty} ^ {+\infty} \int _{-\infty} ^ {+\infty} \int _{-\infty} ^ {+\infty} \int _{-\infty} ^ {+\infty} dK dK' dk dk' \mathcal{I} f(k,k')) .
\end{equation}
 Here $\mathcal{I}$ is given by 
 \begin{align*}
 \resizebox{.97\textwidth}{!}{$
 [ \alpha^* (k',K') \alpha (k,K) b_L ^*(K') b_L(K) - \alpha (k,K) \beta (k',K') b_L(K) b_L(K') +    \alpha (k,K) \alpha (k',K')   b_L(K) b_R ^* (K')  - $} \\ \resizebox{.97\textwidth}{!}{$\alpha (k,K) \beta ^* (k',K')  b_L(K) b_R(K') -  \beta ^* (k,K) \alpha ^* (k',K') b_L ^* (K) b_L ^*(K') + \beta ^* (k,K) \beta (k',K') b_L ^* (K) b_L(K') -$ }\\  \resizebox{.97\textwidth}{!}{$  \beta ^* (k,K) \alpha (k',K') b_L ^* (K) b_R(K')^* + \beta ^* (k,K) \beta ^* (k',K') b_L ^* (K) b_R(K') +   \alpha^* (k',K') \alpha ^* (k,K) b_L ^*(K') b_R(K) - $}\\  \resizebox{.97\textwidth}{!}{$ \alpha (k,K) ^* \beta (k',K') b_R(K) b_L(K') +   \alpha^* (k,K) \alpha (k',K') b_R (K) b_R ^*(K') - \alpha^* (k,K) \beta ^* (k',K') b_R (K) b_R(K') - $ } \\  \resizebox{.97\textwidth}{!}{$ \beta (k,K) \alpha ^* (k',K') b_R ^* (K) b_L ^*(K') +\beta (k,K) \beta (k',K') b_R ^* (K) b_L(K') - \beta (k,K) \alpha (k',K') b_R ^* (K) b_R ^*(K') +$ }\\ \beta (k,K) \beta ^* (k',K')  b_R ^* (K) b_R(K') ],
 \end{align*}
\\
where $\alpha()$ and $\beta ()$ are Bogoliuobov coefficients, and we denote all inertial frame quantities by small letters while Rindler frame quantities by capital letters. One can write the  Bogoliuobov coefficients as (see Appendix of reference \cite{Falcone:2022jbj}, and \cite{PhysRevD.106.045013})
\begin{eqnarray}
    \alpha(k,K) &=& \theta(kK) \sqrt{\frac{K}{k}} G(k,K);  \beta(k,K) = \theta(kK) \sqrt{\frac{K}{k}} G(-k,K);   \\
    G(k,K) &=& \frac{1}{2\pi a} \Gamma \bigg(-\frac{iK}{a} \bigg) \exp \bigg(i \frac{K}{a}\ln{\frac{|k|}{a}} + \sign{(k)} \frac{\pi K}{2a}    \bigg)
\end{eqnarray}

Let us denote the first term inside the exponential in Eq.\eqref{eq:44} as $R_1$ and similarly other terms as $R_2, R_3,......,R_{16}.$  Now, we have to compute these 16 integrations. Let us start with first and perform k,k' integration first.
Say, 
\vspace{-5pt}
\begin{align*}
 \chi _1 (K,K') & =  \int _{-\infty} ^{+\infty} \int _{-\infty} ^{+\infty} dk dk' \alpha ^*(k',K') \alpha (k,K) f(k,k') \\
& = \frac{\sqrt{|K||K'|}}{4 \pi ^2 a^2} \Gamma(\frac{iK'}{a}) \Gamma(\frac{-iK}{a}) \int _{-\infty} ^{+\infty} \int _{-\infty} ^{+\infty} dk dk' \frac{\theta (k'K') \theta(kK))}{\sqrt{|k||k'|}}  \exp \bigg( \frac{-iK'}{a} \ln\frac{|k'|}{a} \\  & + \frac{iK}{a} \ln\frac{|k|}{a} + \sign(k') \frac{\pi K'}{2a} + \sign(k) \frac{\pi K}{2a} \bigg) f(k,k').
\end{align*}
One can express $\theta(kK)$ and $\theta(k'K')$ in terms of sign function and split the integration from $-\infty $  to 0 plus 0 to $ +\infty$. We get the following after changing the limit from 0 to infinity and using properties Eq \eqref{eq:5} to \eqref{eq:8}.
\begin{align*}
\chi_1 (K,K') & = \frac{\sqrt{|K||K'|}}{16 \pi ^2 a^2} \Gamma(\frac{iK'}{a}) \Gamma(\frac{-iK}{a}) [(1-\sign(K'))(1- \sign(K)) e^{\frac{-\pi (K+K')}{2a}} g_4 (K,K')  \\ & +  (1+\sign(K'))(1-\sign(K)) e^{\frac{\pi (-K+K')}{2a}} g_3 (K,K') +  (1-\sign(K'))(1+ \sign(K)) \times \\  & e^{\frac{\pi (K-K')}{2a}} g_2 (K,K')] +  (1+\sign(K'))(1+ \sign(K)) e^{\frac{\pi (K+K')}{2a}} g_1 (K,K')] \delta(K-K').
\end{align*}
\begin{equation} \nonumber
\begin{split}
   \implies R_{1} & =\left[\frac{-1}{2 \pi^{2}} \iint  _{-\infty}^{+\infty} d K d K^{\prime} \chi _{1}\left(K, K^{\prime}\right) b_{L}(K) b_{L}^{*}\left(K^{\prime}\right)\right] \\
    & =\frac{-1}{8 \pi^{4} a^{2}} \int_{-\infty}^{+\infty} d K|K| \left| \Gamma (\frac{ i K }{a} ) \right|^{2}\left[\theta(-K) g_{4}(K) e^{-\pi K / a}+ \theta(K) g_{1}(K)  e^{\pi K/ a}\right]| b_L(K)|^2 
\end{split} 
\end{equation}
Here g(K,K) is denoted by g(K). We compute others in a similar manner. These are given by,

\begin{align*}
 \chi_{2}\left(K, K^{\prime}\right) &=\iint_{-\infty}^{+\infty} d k d k^{\prime} \beta\left(k^{\prime}, K^{\prime}\right) \alpha(k, K) f\left(k, k^{\prime}\right) \\
 \Rightarrow R_{2} &=\left[\frac{1}{2 \pi^{2}} \iint_{-\infty}^{\infty} d K d K^{\prime} \chi _{2}\left(K,K^{\prime}\right) b_{L}(K) b_{L}\left(K^{\prime}\right)\right] \\
&=\frac{1}{8 \pi ^4 a^{2}} \int_{-\infty}^{+\infty} d K|K|\left|\Gamma{\frac{i K}{a}} \right|^2\left[\theta(K) g_{2}+\theta(-K) g_{3}\right] b_{L}(K) b_L (-K) \\
\\
\chi_{3}\left(K, K^{\prime}\right)&=\iint_{-\infty}^{+\infty} d k d k^{\prime} \alpha \left(k^{\prime}, K^{\prime}\right) \alpha(k, K) f\left(k, k^{\prime}\right)\\
\\
 \Rightarrow R_{3} &=\frac{-1}{2 \pi^{2}} \iint_{-\infty}^{\infty} d K d K^{\prime} \chi _{3}\left(K,K^{\prime}\right) b_{L}(K) b_{R} ^* \left(K^{\prime}\right)\\
&= \resizebox{.85 \textwidth}{!}{$ \frac{-1}{8 \pi ^4 a^{2}} \int_{-\infty}^{+\infty} d K|K|\left|\Gamma{\frac{i K}{a}}\right|^{2}\left[\theta(K) g_{2} (K) e^{ \frac{\pi K}{a}} +\theta(-K) g_{3} (K) e^{- \frac{\pi K}{a}}\right] b_{R} ^*(-K)b_L (K) $ } \\
\\
\chi_{4}\left(K, K^{\prime}\right) &=\iint_{-\infty}^{+\infty} d k d k^{\prime} \beta ^* \left(k^{\prime}, K^{\prime}\right) \alpha(k, K) f\left(k, k^{\prime}\right) \\
\Rightarrow R_{4} &=\frac{1}{2 \pi^{2}} \iint_{-\infty}^{\infty} d K d K^{\prime} \chi _{4} \left(K,K^{\prime}\right) b_{L}(K) b_{R} \left(K^{\prime}\right)\\
&=\frac{1}{8 \pi ^4 a^{2}} \int_{-\infty}^{+\infty} d K|K|\left|\Gamma{\frac{i K}{a}} \right|^{2}\left[\theta(-K) g_{4}+\theta(K) g_{1}\right] b_{R} (K) b_L (K) 
\end{align*}
\\
\begin{align*}
 \chi_{5}\left(K, K^{\prime}\right) &=\iint_{-\infty}^{+\infty} d k d k^{\prime} \alpha ^* \left(k^{\prime}, K^{\prime}\right) \beta ^* (k, K) f\left(k, k^{\prime}\right)\\
\\
\Rightarrow R_{5}& =\frac{1}{2 \pi^{2}} \iint_{-\infty}^{\infty} d K d K^{\prime} \chi _{5} \left(K,K^{\prime}\right) b_{L} ^*(K) b_{L} ^*\left(K^{\prime}\right)\\
& =\frac{1}{8 \pi ^4 a^{2}} \int_{-\infty}^{+\infty} d K|K|\left|\Gamma{\frac{i K}{a}} \right|^{2}\left[\theta(-K) g_{2}+\theta(K) g_{3}\right] b_{L} ^* (K) b_L ^* (-K) \\
\\
\chi_{6}\left(K, K^{\prime}\right)& =\iint_{-\infty}^{+\infty} d k d k^{\prime} \beta  \left(k^{\prime}, K^{\prime}\right) \beta ^* (k, K) f\left(k, k^{\prime}\right) \\
\\
 \Rightarrow R_{6}&=\frac{-1}{2 \pi^{2}} \iint_{-\infty}^{\infty} d K d K^{\prime} \chi _{6} \left(K,K^{\prime}\right) b_{L} ^*(K) b_{L} \left(K^{\prime}\right)\\
& =\frac{-1}{8 \pi ^4 a^{2}} \int_{-\infty}^{+\infty} d K|K|\left|\Gamma{\frac{i K}{a}} \right|^{2}\left[\theta(K) e^{- \pi K/a} g_{4}+\theta(-K) e^{ \pi K/a}g_{1}\right] b_{L} (-K) b_L ^* (-K) 
\\
\\
 \chi_{7}\left(K, K^{\prime}\right)&=\iint_{-\infty}^{+\infty} d k d k^{\prime} \alpha  \left(k^{\prime}, K^{\prime}\right) \beta ^* (k, K) f\left(k, k^{\prime}\right) \\
\\
 \Rightarrow R_{7} &=\frac{1}{2 \pi^{2}} \iint_{-\infty}^{\infty} d K d K^{\prime} \chi _{7} \left(K,K^{\prime}\right) b_{L} ^*(K) b_{R} ^* \left(K^{\prime}\right)\\
&=\frac{1}{8 \pi ^4 a^{2}} \int_{-\infty}^{+\infty} d K|K|\left|\Gamma{\frac{i K}{a}} \right|^{2}\left[\theta(K) g_{4}+\theta(-K) g_{1}\right] b_{L} (-K) ^* b_R ^* (-K) \\
\\
 \chi_{8}\left(K, K^{\prime}\right)&=\iint_{-\infty}^{+\infty} d k d k^{\prime} \beta ^*  \left(k^{\prime}, K^{\prime}\right) \beta ^* (k, K) f\left(k, k^{\prime}\right) \\
\\
\Rightarrow R_{8}&=\frac{-1}{2 \pi^{2}} \iint_{-\infty}^{\infty} d K d K^{\prime} \chi _{8} \left(K,K^{\prime}\right) b_{L} ^*(K) b_{R} \left(K^{\prime}\right)\\
&=\frac{-1}{8 \pi ^4 a^{2}} \int_{-\infty}^{+\infty} d K|K|\left|\Gamma{\frac{i K}{a}} \right|^{2}\left[\theta(-K) e^{ \pi K/a} g_{2}+\theta(K) e^{- \pi K/a}g_{3}\right] b_{L} ^* (-K)  b_R  (K)
\end{align*}
\\ 
\begin{align*}
\chi_{9}\left(K, K^{\prime}\right) &=\iint_{-\infty}^{+\infty} d k d k^{\prime} \alpha ^*  \left(k^{\prime}, K^{\prime}\right) \alpha ^* (k, K) f\left(k, k^{\prime}\right) \\
\Rightarrow R_{9} &=\frac{-1}{2 \pi^{2}} \iint_{-\infty}^{\infty} d K d K^{\prime} \chi _{9} \left(K,K^{\prime}\right) b_{R} (K) b_{L} ^* \left(K^{\prime}\right)\\
&=\frac{-1}{8 \pi ^4 a^{2}} \int_{-\infty}^{+\infty} d K|K|\left|\Gamma{\frac{i K}{a}} \right|^{2}\left[\theta(-K) e^{ - \pi K/a} g_{2}+\theta(K) e^{ \pi K/a}g_{3}\right] b_{L} ^* (K)  b_R  (-K)\\
\\
\chi_{10}\left(K, K^{\prime}\right)&=\iint_{-\infty}^{+\infty} d k d k^{\prime} \beta  \left(k^{\prime}, K^{\prime}\right) \alpha ^* (k, K) f\left(k, k^{\prime}\right) \\
 \Rightarrow R_{10}&=\frac{1}{2 \pi^{2}} \iint_{-\infty}^{\infty} d K d K^{\prime} \chi _{10} \left(K,K^{\prime}\right) b_{R} (K) b_{L} \left(K^{\prime}\right)\\
& =\frac{1}{8 \pi ^4 a^{2}} \int_{-\infty}^{+\infty} d K|K|\left|\Gamma{\frac{i K}{a}} \right|^{2}\left[\theta(K)  g_{4}+\theta(-K) g_{1}\right] b_{L} (-K)  b_R  (-K)\\
\\
\chi_{11}\left(K, K^{\prime}\right) &=\iint_{-\infty}^{+\infty} d k d k^{\prime} \alpha  \left(k^{\prime}, K^{\prime}\right) \alpha ^* (k, K) f\left(k, k^{\prime}\right) 
\\
\Rightarrow R_{11}&=\frac{-1}{2 \pi^{2}} \iint_{-\infty}^{\infty} d K d K^{\prime} \chi _{11} \left(K,K^{\prime}\right) b_{R} (K) b_{R} ^* \left(K^{\prime}\right)\\
& =\frac{-1}{8 \pi ^4 a^{2}} \int_{-\infty}^{+\infty} d K|K|\left|\Gamma{\frac{i K}{a}} \right|^{2}\left[\theta(K) e^ {\pi K/a}  g_{4}+\theta(-K) g_{1} e^ {-\pi K/a} \right] b_{R} (-K) ^*   b_R  (-K)\\
\\
 \chi_{12}\left(K, K^{\prime}\right)&=\iint_{-\infty}^{+\infty} d k d k^{\prime} \beta ^*  \left(k^{\prime}, K^{\prime}\right) \alpha ^* (k, K) f\left(k, k^{\prime}\right) 
\\
\Rightarrow R_{12}&=\frac{1}{2 \pi^{2}} \iint_{-\infty}^{\infty} d K d K^{\prime} \chi _{12} \left(K,K^{\prime}\right) b_{R} (K) b_{R} \left(K^{\prime}\right)\\
& =\frac{1}{8 \pi ^4 a^{2}} \int_{-\infty}^{+\infty} d K|K|\left|\Gamma{\frac{i K}{a}} \right|^{2}\left[\theta(K)  g_{3}+\theta(-K) g_{2} \right] b_{R} (-K)  b_R  (K)\\
\\
\chi_{13}\left(K, K^{\prime}\right)&=\iint_{-\infty}^{+\infty} d k d k^{\prime} \alpha \left(k^{\prime}, K^{\prime}\right) \beta (k, K) f\left(k, k^{\prime}\right) \\
\\
\Rightarrow R_{13}& =\frac{1}{2 \pi^{2}} \iint_{-\infty}^{\infty} d K d K^{\prime} \chi _{13} \left(K,K^{\prime}\right) b_{R} ^* (K) b_{L} ^* \left(K^{\prime}\right)\\
& =\frac{1}{8 \pi ^4 a^{2}} \int_{-\infty}^{+\infty} d K|K|\left|\Gamma{\frac{i K}{a}}\right|^{2}\left[\theta(-K)  g_{4}+\theta(K) g_{1} \right] b_{R} ^* (K)  b_L ^* (K)
\end{align*}
\\
\begin{align*}
\chi_{14}\left(K, K^{\prime}\right)&=\iint_{-\infty}^{+\infty} d k d k^{\prime} \beta \left(k^{\prime}, K^{\prime}\right) \beta (k, K) f\left(k, k^{\prime}\right) \\
\Rightarrow R_{14}& =\frac{-1}{2 \pi^{2}} \iint_{-\infty}^{\infty} d K d K^{\prime} \chi _{14} \left(K,K^{\prime}\right) b_{R} ^* (K) b_{L} \left(K^{\prime}\right)\\
& =\frac{-1}{8 \pi ^4 a^{2}} \int_{-\infty}^{+\infty} d K|K|\left|\Gamma{\frac{i K}{a}}\right|^{2}\left[\theta(-K) e^{\pi K/a}  g_{3}+\theta(K) e^{-\pi K/a} g_{2} \right] b_{R} ^* (K)  b_L (-K)\\
\\
\chi_{15}\left(K, K^{\prime}\right)&=\iint_{-\infty}^{+\infty} d k d k^{\prime} \alpha \left(k^{\prime}, K^{\prime}\right) \beta (k, K) f\left(k, k^{\prime}\right) \\
\Rightarrow R_{15}&=\frac{1}{2 \pi^{2}} \iint_{-\infty}^{\infty} d K d K^{\prime} \chi _{15} \left(K,K^{\prime}\right) b_{R} ^* (K) b_{R} ^* \left(K^{\prime}\right)\\
& =\frac{1}{8 \pi ^4 a^{2}} \int_{-\infty}^{+\infty} d K|K|\left|\Gamma{\frac{i K}{a}} \right|^{2}\left[\theta(K) g_{2}+\theta(-K) g_{3} \right] b_{R} ^* (-K)  b_R ^* (K)\\
\\
\chi_{16}\left(K, K^{\prime}\right)&=\iint_{-\infty}^{+\infty} d k d k^{\prime} \beta ^* \left(k^{\prime}, K^{\prime}\right) \beta (k, K) f\left(k, k^{\prime}\right) \\
\Rightarrow R_{16}&=\frac{-1}{2 \pi^{2}} \iint_{-\infty}^{\infty} d K d K^{\prime} \chi _{16} \left(K,K^{\prime}\right) b_{R} ^* (K) b_{R} \left(K^{\prime}\right)\\
& =\frac{-1}{8 \pi ^4 a^{2}} \int_{-\infty}^{+\infty} d K|K|\left|\Gamma{\frac{i K}{a}} \right|^{2}\left[\theta(-K) e^{\pi K/a} g_{4}+\theta(K) e^{-\pi K/a}  g_{1} \right] b_{R} ^* (K)  b_R (K)
\end{align*}

Now, we put all these together in equation \eqref{eq:16} and separate terms containing $b_L(K)$, as our aim is to first take trace over $b_L(K)$. 
\begin{align} \label{eq:45}
W_{R} &=\resizebox{.96\textwidth}{!}{$ \bar{N} \exp \Biggl[\frac{-1}{8 \pi^{4} a^{2}} \int_{-\infty}^{+\infty} d K \mid K \mid \left|\Gamma \left(\frac{i K}{a}\right)\right|^{2} \biggl[ \left( \theta(-K) g_{4}(K) e^{-\frac{\pi K }{a}} + \theta(K) g_{1}(K) e^{\frac{\pi K}{a}} \right) 
\left|b_{L}(K)\right|^{2} $} \notag \\  
&\quad + b_{L}(K) \Bigl\{-\left(\theta(K) g_{2} + \theta(-K) g_{3}\right) b_{L}(-K) + \left(\theta(K) e^{\frac{\pi K }{a}} g_{2} + \theta(-K) e^{-\frac{\pi K}{a}} g_{3}\right) b_{R}^{*}(-K) \notag \\ 
&\quad -\left(\theta(-K) g_{4} + \theta(K) g_{1}\right) b_{R}(K) \Bigr\}  
+ b_{L}^{*}(K) \Bigl\{-\left(\theta(-K) g_{2} + \theta(K) g_{3}\right) b_{L}^{*} (-K) \notag \\ 
&\quad + \left(\theta(K) e^{\frac{\pi K }{a}} g_{3} + \theta(-K) e^{-\frac{\pi K}{a}} g_{2}\right) b_{R}(-K) - \left(\theta(-K) g_{4} + \theta(K) g_{1}\right) b_{R}^{*} (K) \Bigr\} \notag \\ 
&\quad + \left( \theta(K) e^{-\frac{\pi K }{a}} g_{4} + \theta(-K) e^{\frac{\pi K}{a}} g_{1}\right) \left|b_{L}(-K)\right|^{2} 
+ b_{L}(-K) \Bigl\{-\left(\theta(K) g_{4} + \theta(-K) g_{1} \right) b_{R}(-K) \notag \\ 
&\quad + \left(\theta(-K) e^{\pi K/a} g_{3} + \theta(K) g_{2} e^{-\pi K/a}\right) b_{R}^{*}(K)\Bigr\} \notag \\ 
&\quad + b_{L}^{*}(-K) \Bigl\{\left(\theta(-K) e^{\pi K/a} g_{2} + \theta(K) g_{3} e^{-\pi K/a}\right) b_{R}(K) - \left(\theta(K) g_{4} + \theta(-K) g_{1}\right) b_{R}^{*}(-K) \Bigr\} \notag \\ 
&\quad + \left(\theta(K) e^{\pi K/a} g_{4} + \theta(-K) e^{-\pi K/a} g_{1}\right) \left|b_{R}(-K)\right|^{2} 
- \left(\theta(K) g_{3} + \theta(-K) g_{2}\right) b_{R}(K) b_{R}(-K) \notag \\ 
&\quad - \left(\theta(K) g_{2} + \theta(-K) g_{3}\right) b_{R}^{*}(-K) b_{R}^{*}(K) 
+ \left(\theta(-K) e^{\pi K/a} g_{4} + \theta(K) e^{-\pi K/a} g_{1}\right) \left|b_{R}(K)\right|^{2} \biggl] \Biggr]\\
&= e^{w} \bar{N} \exp \left[\int _0 ^\infty d K \left\{-\tilde{M}_{1}\left|b_{L}(K)\right|^{2} + b_{L}(K) \tilde{M}_{2} + b_{L}^{*}(K) \tilde{M}_{3} - \tilde{M}_{4}\right\}\right] \numberthis \label{eq:46}
\end{align}
\\
where we have denoted terms that do not contain $b_L(K)$ by w, $\tilde{M_1}$, $\tilde{M_2}$, $\tilde{M_3}$ and $\tilde{M_4 }$ , and they are given by following expressions.\\

\begin{align*} \numberthis \label{eq:47}
\tilde{M_1} & = \frac{1}{8 \pi^{4} a^2 } \mid K \mid \left|\Gamma {\frac{i K}{a}}\right|^{2} [ g_1 (K) e^{\pi K/a} + g_1(-K) e^{-\pi K/a} ] \\
w &= \frac{-1}{8 \pi^{4} a^2 } \int_{-\infty}^{\infty} dK \mid K \mid \left|\Gamma \left(\frac{i K}{a}\right)\right|^{2} \biggl[ \left(\theta (K) g_4 (K) e^{\pi K/a}  + \theta (-K) g_1 (K) e^{-\pi K/a}\right) \left|b_R(-K)\right|^2 \notag \\ 
&\quad - \left(\theta (K) g_3 (K) + \theta (-K) g_2 (K)\right) b_R (K) b_R (-K) - \left(\theta (K) g_2 (K) + \theta (-K) g_3 (K)\right) b_R^* (-K) b_R^*(K) \notag \\ 
&\quad + \left(\theta (-K) g_4 (K) e^{\frac{\pi K}{a}} + \theta (K) g_1 (K) e^{\frac{-\pi K}{a}}\right) \left|b_R (K)\right|^2 \biggl] \\
\tilde{M_2} & = \frac{-1}{8 \pi^{4} a^2 } \mid K \mid \left|\Gamma {\frac{i K}{a}}\right|^{2} [-g_2 (K) b_L (-K) + g_2 (K) e^{\pi K/a} b_R ^* (-K) - g_1 (K) b_R (K) \\ & - g_1 (-K) b_R (K) +  g_3(-K) e^{-\pi K/a} b_R ^* (-K) - g_3(-K) b_L (-K) ]; \\
\tilde{M_3} & = \frac{-1}{8 \pi^{4} a^2 } \mid K \mid \left|\Gamma {\frac{i K}{a}}\right|^{2} [-g_3 (K) b_L ^* (-K) + g_3 (K) e^{\pi K/a} b_R (-K) -  g_1 (K) b_R ^* (K) -  \\ & g_1 (-K) b_R ^* (K) + 
 g_2(-K) e^{-\pi K/a} b_R (-K) - g_2(-K) b_L ^* (-K) ] \\ 
\tilde{M_4}  & = \frac{1}{8 \pi^{4} a^2 } \mid K \mid \left|\Gamma {\frac{i K}{a}}\right|^{2} [ |b_L (-K)|^2 ( g_4 (-K) e^ {\frac{\pi K}{a}} + g_4 (K) e^ {\frac{-\pi K}{a}} ) + b_L (-K) ( g_3 (-K) e^ { \frac{\pi K}{a}} b_R ^* (K) \\ & - g_4 (-K) b_R (-K) - g_4(K) b_R (-K) + g_2 (K) e^ {-\pi K/a} b_R ^* (K) )  + b_L^* (-K) ( g_2 (-K) e^{\pi K/a} b_R (K)  \\ &
-g_4 (-K) b_R ^* (-K)+ g_3 (K) e^{- \pi K/a} b_R (K) - g_4 (K) b_R ^* (-K) ) ]
\end{align*}
\\
\\
Since $b_L(K)$ is complex we can let $b_L(K) = x+i y \Rightarrow b_{L}^{*}(K)=x-i y $,  where x and y are real. Therefore,
\\
\begin{align*}
W_R & =\bar{N} e^{w} \exp \left[\int d K \left\{ -\left(x^{2}+y^{2}\right) \tilde{M} _{1} +(x+i y) \tilde{M}_{2}+(x-i y) \widetilde{M_{3}}-\tilde{M}_{4}\right\} \right] \\
 & =\bar{N} e^{w} \exp \left[\int d K\left\{-x^{2} \tilde{M}_{1}-y^{2} \tilde{M}_{1}+x\left(\tilde{M}_{2}+\tilde{M}_{3}\right)+y i\left(\tilde{M}_{2}-\tilde{M}_{3}\right)-\tilde{M}_{4}\right\} \right]
\end{align*}
Taking trace over $b_L$(K) we have \\
\begin{align*}
& \bar{N} e^{w} \iint _{-\infty} ^{\infty} dx dy \exp\left[-x^{T} \tilde{M}_{1} x+\left(\tilde{M}_{2}+\tilde{M}_{3}\right) x-y^{T} \tilde{M}_{1} y+ i y \left( \tilde{M_{2}}-\tilde{M}_{3}\right)-\tilde{M}_{4}\right] \\ &
= \bar{N}^{\prime} e^{w} \exp \left[\frac{1}{4}\left(\tilde{M}_{2}+\tilde{M}_{3}\right) \tilde{M}_{1}^{-1}\left(\tilde{M}_{2}+\tilde{M}_{3}\right)-\frac{1}{4}\left(\tilde{M}_{2} - \tilde{M}_{3}\right) \tilde{M}_{1}^{-1}\left(\tilde{M}_{2} - \tilde{M}_{3}\right)-\tilde{M}_{4}\right]
\end{align*}
Where $\bar{N}^{\prime}$ is the new normalization factor. We substitute $\tilde{M}s $ and get the expression that contains $b_L(-K)$. We repeat the similar work to trace over $b_L (-K)$ and finally we get the following reduced Wigner distribution for the right Rindler wedge.
    
\begin{multline} \label{eq:48}
    W_{R_{Reduced}} = N \exp \Bigl[ \frac{-1}{8 \pi ^4 a ^2} \int _0 ^{\infty} dK \mid K \mid \left| \Gamma \frac{iK}{a} \right| ^2 [J(K) | b_R (K) | ^2 + R_1 b_R ^* (K) b_R ^* (-K) \\
    + R_2 b_R(K) b_R (-K) + L |b_R (-K)|^2 \Bigr]
\end{multline}
where $J(K),R_1(K),R_2(K)$ and $L(K)$ are give by following expressions
\begin{align*}
J(K) = - \left( \frac{-(g_1 (K) + g_1 (-K)) (g_2 (K) + g_3 (-K))}{g_1 (K) e^{\pi K/a} + g_1 (-K) e^{-\pi K/a}} + (e^{\pi K/a} g_3 (-K) + g_2 (K) e^{-\pi K/a}) \right) \times \\
\\
\left( \frac{ - (g_3 (K) + g_2 (-K)) (g_1(K) + g_1 (-K))}{g_1 (K) e^{\pi K/a} + g_1 (-K) e^{- \pi K/a}} + (g_2 (-K) e^{\pi K/a} + g_3(K) e^{-\pi K/a}) \right) \times \\
 \left\{ - \frac{(g_2 (K) + g_3 (-K)) (g_2 (-K) + g_3 (K)) }{g_1 (K) e^{\pi K/a} + g_1 (-K) e^{- \pi K/a}}+ (g_4 (-K) e^{\pi K/a} + g_4 (K) e^{-\pi K/a}) \right\} ^{-1} \\ 
- \frac{(g_1 (K) + g_1 (-K))^2}{g_1 (K) e^{\pi K/a } + g_1 (-K) e^{-\pi K/a}}
 + (g_1 (K) e^{- \pi K/a }+ g_1 (-K) e^{ \pi K/a }) \numberthis \label{eq:49}\\
\\
R_1= - \left( \frac{-(g_1 (K) + g_1 (-K)) (g_2 (K) + g_3 (-K))}{g_1 (K) e^{\pi K/a} + g_1 (-K) e^{-\pi K/a}} + (e^{\pi K/a} g_3 (-K) + g_2 (K) e^{-\pi K/a}) \right) \times \\
\\
\left( \frac{ (g_3 (K) + g_2 (-K)) (g_2(K) e^{\pi K/a} + g_3 (-K) e^{- \pi K/a})}{g_1 (K) e^{\pi K/a} + g_1 (-K) e^{- \pi K/a}} + (-g_4 (-K)  - g_4(K) ) \right) \times \\
\left\{ - \frac{(g_2 (K) + g_3 (-K)) (g_2 (-K) + g_3 (K)) }{g_1 (K) e^{\pi K/a} + g_1 (-K) e^{- \pi K/a}}+ (g_4 (-K) e^{\pi K/a} + g_4 (K) e^{-\pi K/a}) \right\} ^{-1}  \\ +
\frac{(g_1 (K) + g_1 (-K)) (g_2 (K) e^{\pi K/a} + g_3 (-K) e^{-\pi K/a})}{g_1 (K) e^{\pi K/a } + g_1 (-K) e^{-\pi K/a}}
 - (g_2 (K)+ g_3 (-K) )  \numberthis \label{eq:50}\\
R_2 = - \left( \frac{(g_3 (K) e^{\pi K/a}+ g_2 (-K)e^{- \pi K/a}) (g_2 (K) + g_3 (-K))}{g_1 (K) e^{\pi K/a} + g_1 (-K) e^{-\pi K/a}} + (-g_4 (K) - g_4 (-K) ) \right) \times \\
\\
\left( \frac{ - (g_3 (K) + g_2 (-K)) (g_1(K) + g_1 (-K))}{g_1 (K) e^{\pi K/a} + g_1 (-K) e^{- \pi K/a}} + (g_2 (-K) e^{\pi K/a} + g_3(K) e^{-\pi K/a}) \right) \times \\
\left\{ - \frac{(g_2 (K) + g_3 (-K)) (g_2 (-K) + g_3 (K)) }{g_1 (K) e^{\pi K/a} + g_1 (-K) e^{- \pi K/a}}+ (g_4 (-K) e^{\pi K/a} + g_4 (K) e^{-\pi K/a}) \right\} ^{-1} \\ +
\frac{(g_1 (K) + g_1 (-K)) (g_3 (K) e^{\pi K/a} + g_2 (-K) e^{- \pi K/a})}{g_1 (K) e^{\pi K/a } + g_1 (-K) e^{-\pi K/a}} - (g_3 (K) + g_2 (-K) ) \numberthis \label{eq:51}
\end{align*}
\begin{align*}
L(K) = - \left( \frac{(g_3 (K) e^{\pi K/a}+ g_2 (-K)e^{- \pi K/a}) (g_2 (K) + g_3 (-K))}{g_1 (K) e^{\pi K/a} + g_1 (-K) e^{-\pi K/a}} + (-g_4 (K) - g_4 (-K) ) \right) \times \\
\\
\left( \frac{ (g_3 (K) + g_2 (-K)) (g_2(K) e^{\pi K/a} + g_3 (-K) e^{- \pi K/a})}{g_1 (K) e^{\pi K/a} + g_1 (-K) e^{- \pi K/a}} + (-g_4 (-K)  - g_4(K) ) \right) \times \\
\left\{ - \frac{(g_2 (K) + g_3 (-K)) (g_2 (-K) + g_3 (K)) }{g_1 (K) e^{\pi K/a} + g_1 (-K) e^{- \pi K/a}}+ (g_4 (-K) e^{\pi K/a} + g_4 (K) e^{-\pi K/a}) \right\} ^{-1} \\  -
\frac{(g_2 (K) e^{\pi K/a} + g_3 (-K) e^{-\pi K/a}) (g_3 (K) e^{\pi K/a} + g_2 (-K) e^{- \pi K/a})}{g_1 (K) e^{\frac{\pi K}{a} } + g_1 (-K) e^{\frac{-\pi K}{a}}} + \\
(g_4 (K) e^{\frac{\pi K}{a}} + g_4 (-K) e^{\frac{-\pi K}{a}} ) \numberthis \label{eq:52}
\end{align*}
\section{\bf{ Calculating $\langle n_M(k) \rangle$ for section 3.3}} \label{Appendix B}
The two point function $f(k,k')$, discussed in section 3.3 is given by the following expression.
\begin{equation} \label{eq:53}
    f(k,k') = \sqrt{\frac{4\pi}{\gamma}} \frac{\theta(kk')}{\sqrt{|k||k'|}} e^{-(\ln(k'/k))^2/\gamma}
\end{equation}
To obtain the expectation value of the number density in an inertial frame, one can use the following definition of the inverse function $f^{-1} (k',k'')$
\begin{equation} \label{eq:54}
     \int f(k,k')f^{-1} (k',k'') dk'  = \delta(k-k'').
\end{equation} 
The above expression, denoted by Eq.\eqref{eq:53} represents Dirac delta in the limit $\gamma \rightarrow 0 $. Therefore, to get the inverse for a small deviation of $f(k,k')$ from the identity, $\delta(k-k'')$, we use the analogy with the matrix form of $f(k,k')$. Using the identity for infinite-dimensional matrices, $A^{-1}=I + (I-A) + (I-A)^2 +....$ for $||I-A||< 1$, we obtain
\begin{align} \label{eq:55}
    \left(\frac{f(k,k')}{2 \pi}\right)^{-1}  = & 2 \delta (k-k') - \frac{k'\theta (kk') \delta (k-k')}{\sqrt{kk'}} - \frac{\gamma \theta(kk') \delta'' (\log{\frac{k}{k'}})}{4 \sqrt{k k'}} - \mathcal{O} (\gamma ^2) \\
    =& \delta (k-k') - \frac{\gamma \theta(kk') \delta'' (\log{\frac{k}{k'}})}{4 \sqrt{k k'}} - \mathcal{O} (\gamma ^2) \label{eq:56}
\end{align}
One can verify the aforementioned result to the first order in $\gamma $ as follows:
\begin{align*}
    \int f(k,k')f^{-1} (k',k'') dk'   = & \int dk'\bigg(  \frac{\theta(kk') k \delta(k-k')  }{\sqrt{|k||k'|}} + \gamma \frac{\theta(kk')}{4 \sqrt{|k||k'|}} \delta''(\ln{(k/k')}) +.... \bigg) \times \\ &  \bigg( 2  \delta(k'-k'') -  \frac{\theta(k'k'') k'' \delta(k'-k'')  }{\sqrt{|k'||k''|}}  - \frac{\gamma  \theta(k'k'')}{4 \sqrt{|k'||k''|}} \delta''(\ln{(k'/k'')}) \\
   & + ...   \bigg)  \\
    = & \int dk' \bigg( \frac{\theta(kk') k' \delta(k-k')  }{\sqrt{|k||k'|}} 2 \delta(k' - k'')   \\ & + \gamma \frac{\theta(kk')}{2 \sqrt{|k||k'|}} \delta''(\ln{(k/k')}) \delta(k' - k'')  \\ & -
   \frac{\theta(kk') \theta(k'k'') k'k'' \delta(k-k') \delta(k'-k'')  }{\sqrt{|k||k'||k'||k''|}}  \\ & - \frac{\gamma \theta(kk') \theta(k'k'') k'' \delta(k'-k'') \delta''(\ln{(k/k')})  }{4 \sqrt{|k||k'||k'||k''|}}\\ & - \frac{\gamma \theta(kk') \theta(k'k'') k' \delta(k-k') \delta''(\ln{(k'/k'')})  }{4 \sqrt{|k||k'||k'||k''|}} - \mathcal{O}(\gamma^2)\bigg) \\
   =& \frac{2 \theta(kk'') k'' \delta(k-k'')  }{\sqrt{|k||k''|}} - \frac{ \theta(kk'') k'' \delta(k-k'')  }{\sqrt{|k||k''|}} \\
   = & \delta(k-k'') \numberthis
\end{align*}
Here, the Taylor series expansion of \( f(k,k') \) about \(\gamma = 0\) is employed in the first line, and in the final step, \(\theta(kk') = 1\) is used, since it is multiplied by the Dirac delta, which peaks at \( k = k' \). The prime over the Dirac delta denotes the derivative with respect to $\ln{(k/k')}$.  Substituting the inverse from Eq.\eqref{eq:56} in equation \eqref{eq:13}, we get the following expectation value of number density in an inertial frame.
\begin{equation} \label{eq:57}
    \langle n_M(k) \rangle = -\bigg(\frac{\gamma \delta'' (\log{k/k'})}{8 \sqrt{kk'}} + \mathcal{O}(\gamma ^2) \bigg)_{k=k'}
\end{equation}
\\
Considering the number density expectation value in the above equation as a distribution under integration, we use \( \int g(x)\delta''(x) = \int g''(x) \delta(x) \) \cite{Cortizo:1995rj} to formally write the expectation value as
\begin{equation} \label{eq:58}
    \langle n_M(k) \rangle = -\frac{\gamma \delta(0)}{32} + \mathcal{O}(\gamma ^2).
\end{equation}

\end{appendices}

\pagebreak

\bibliographystyle{./style_files/JHEP}
\bibliography{references}



\end{document}

%% file: style_files/packages.tex
\usepackage{braket}
\usepackage{color}
\usepackage[dvipsnames]{xcolor}

\newcommand{\be}{\begin{equation}}
\newcommand{\ee}{\end{equation}}
\newcommand{\ba}{\begin{eqnarray}}
\newcommand{\ea}{\end{eqnarray}}

\usepackage{colonequals}

\usepackage{ulem}

%% file: style_files/PSI_packages.tex
\usepackage[utf8]{inputenc}
\usepackage[T1]{fontenc}
\usepackage{fancyhdr}
\usepackage{lipsum}
\usepackage{graphicx}
\usepackage{titlesec}
\usepackage{appendix}
\usepackage[sectionbib]{chapterbib}
\usepackage[breakwords]{truncate}
\usepackage{lastpage}
\usepackage{amsmath}
\usepackage{hyperref}
\usepackage{amssymb}
\usepackage[font={small}]{caption}
\usepackage{physics}
